\documentclass{aa}

\usepackage{graphicx}
\usepackage{multirow}
\usepackage{subcaption}
\usepackage{listings}
\usepackage{adjustbox}
\usepackage{booktabs}
\usepackage{xcolor}
\usepackage{soul}
\usepackage{caption}
\usepackage{float}
\usepackage[normalem]{ulem}

\usepackage[
  text={Reproduced with permission from Astronomy \& Astrophysics, © ESO},
  angle=0,
  vpos=285mm,
  fontsize=14pt,
  color=black,
  firstpage        
]{draftwatermark}

\usepackage{txfonts}
\usepackage{hyperref}
\usepackage{url}
\newcommand{\orcid}[1]{\href{https://orcid.org/#1}{\textcolor[HTML]{A6CE39}{\aiOrcid}}}
\usepackage{orcidlink}

\definecolor{codegreen}{rgb}{0,0.6,0}
\definecolor{codegray}{rgb}{0.5,0.5,0.5}
\definecolor{codepurple}{rgb}{0.58,0,0.82}
\definecolor{backcolour}{rgb}{0.95,0.95,0.92}

\lstdefinestyle{mystyle}{
    backgroundcolor=\color{backcolour},   
    commentstyle=\color{codegreen},
    keywordstyle=\color{blue},
    numberstyle=\tiny\color{codegray},
    stringstyle=\color{red},
    basicstyle=\ttfamily\footnotesize,
    breakatwhitespace=false,         
    breaklines=true,                 
    captionpos=b,                    
    keepspaces=true,                 
    numbers=none,                    
    numbersep=5pt,                  
    showspaces=false,                
    showstringspaces=false,
    showtabs=false,                  
    tabsize=4,
    escapeinside={<@}{@>}
}
\lstdefinelanguage{Python}
{
  morekeywords={from, import, def, return},
  comment=[l]{\#},
  morestring=[b]",
  alsodigit={-},
  alsoletter={&},
  moredelim=[is][\color{brown}\ttfamily]{[}{]}
}

\lstdefinestyle{pythonstyle}{
    language=Python,
    basicstyle=\ttfamily\small,
    keywordstyle=\color{purple},
    stringstyle=\color{red},
    breaklines=true,
    showstringspaces=false
}

\makeatletter
\g@addto@macro\appendix{
  \counterwithin*{lstlisting}{section}
  \renewcommand{\thelstlisting}{\thesection.\arabic{lstlisting}}
  \setcounter{lstlisting}{0}
}
\makeatother

\begin{document}

   \title{Querying an astronomical database using large language models: the ALeRCE text-to-SQL system}

   \author{P.A. Estévez \inst{1,2} \orcidlink{0000-0001-9164-4722} \and J. Espejo-Moreira \inst{1,2} \orcidlink{0009-0004-2094-7303} \and S. Sanfeliú-Alvarez \inst{1, 5} \and F. Förster \inst{2,3,4,5} \orcidlink{0000-0003-3459-2270} \and A. M. Muñoz Arancibia \inst{2,4} \orcidlink{0000-0002-8722-516X} \and G. Cabrera-Vives \inst{6,2,7,8} \orcidlink{0000-0002-2720-7218} \and F. E. Bauer \inst{9} \orcidlink{0000-0002-8686-8737}  \and A. Bayo \inst{10} \orcidlink{0000-0001-7868-7031}   \and M. Catelan \inst{2,11,12} \orcidlink{0000-0001-6003-8877}  \and R. Dastidar \inst{2} \orcidlink{0000-0001-6191-7160} \and L. Hernández-García \inst{13,14,2} \orcidlink{0000-0002-8606-6961} \and J.A. Intriago\inst{1} \orcidlink{0000-0001-8964-7385} \and G. Pignata \inst{9} \orcidlink{0000-0003-0006-0188}
    }

   \institute{Department of Electrical Engineering, University of Chile, Av. Tupper 2007, Santiago, Chile \and
   Millennium Institute of Astrophysics (MAS), Nuncio Monseñor Sótero Sanz 100, Providencia, Santiago, Chile \and Data and Artificial Intelligence Initiative (ID\&IA), Universidad de Chile \and Center for Mathematical Modeling, Universidad de Chile, Beauchef 851, North building, 7th floor, Santiago 8320000, Chile \and Departamento de Astronom\'ia, Universidad de Chile, Casilla 36D, Santiago, Chile \and Department of Computer Science, Universidad de Concepción, Edmundo Larenas 219, Concepción, Chile \and Center for Data and Artificial Intelligence, Universidad de Concepción, Edmundo Larenas 310, Concepción, Chile \and Heidelberg Institute for Theoretical Studies, Heidelberg, Baden-Württemberg, Germany \and Instituto de Alta Investigación, Universidad de Tarapacá, Casilla 7D, Arica, 1010000, Chile   \and European Southern Observatory, Karl-Schwarzschild-Strasse 2, 85748 Garching bei München, Germany \and Instituto de Astrofísica, Facultad de Física, Pontificia Universidad Católica de Chile, Casilla 306, Santiago 22, Chile \and Centro de Astroingeniería, Pontificia Universidad Católica de Chile, Av. Vicuña Mackenna 4860, 7820436 Macul, Santiago, Chile  
     \and Instituto de Estudios Astrof\'isicos, Facultad de Ingenier\'ia y Ciencias, Universidad Diego Portales, Av. Ej\'ercito Libertador 441, Santiago, Chile
\and Centro Interdisciplinario de Data Science, Facultad de Ingenier\'ia y Ciencias, Universidad Diego Portales, Av. Ej\'ercito Libertador 441, Santiago, Chile}
   \date{Received November 22, 2025}
 
  \abstract
   {We develop a text-to-SQL (structured query language) system based on large language models (LLMs) using in-context learning and apply it to the Automatic Learning for the Rapid Classification of Events (ALeRCE) astronomical database. ALeRCE is a community broker for the Zwicky Transient Facility and the Vera C. Rubin Observatory. 
   The system enables users to query the database in natural language (NL) and generates executable SQL queries. To develop and evaluate the system, we constructed a dataset of 110 NL/SQL pairs. We propose a step-by-step generation framework comprising four modules: schema linking, query classification, prompt decomposition, and self-correction. The performance of thirteen LLMs is evaluated using in-context learning and prompt engineering techniques. Text-to-SQL performance is assessed using the perfect-match (PM) rate for row identifiers (e.g., object identifiers) and column identifiers (i.e., column names). The proposed step-by-step framework consistently outperforms a direct-inference baseline, while the self-correction module consistently reduces execution errors. For Claude Opus 4.6, PM performance on row (column) identifiers is high for simple queries, reaching 0.97 (0.94), and decreases with query complexity to 0.44 (0.72) for medium queries and 0.59 (0.49) for hard queries. Among the thirteen evaluated models, the best-performing LLMs for the text-to-SQL task are Claude Opus 4.6, Gemini 2.5 Pro, Gemini 3 Flash, and GPT-5.2-Codex.
   }


   \maketitle

\section{Introduction}
\label{sec:introduction}
Large Language Models (LLMs) are revolutionizing the field of natural language processing (NLP) and its applications. These models are pre-trained on massive amounts of text data and can be fine-tuned to specific tasks such as language translation or question answering. In the field of astronomy, the main applications up to now have been on the named entity recognition (NER) task \citep{grezes2021building,ghosh-etal-2022-astro,sotnikov2023language, Shao_2024}, question-answering \citep{perkowski2024astrollama},  hypothesis generation \citep{ciucua2023harnessing}, text summarization \citep{platform_astronomy}, and Artificial Intelligence (AI) research assistants \citep{joseph2026astrovisbench, ye2025replicationbench}.
NER aims to identify and classify named entities in text, such as organizations, citations, surveys, telescopes, and celestial objects. For example, AstroBERT \citep{grezes2021building} identifies organizations in the acknowledgment section of papers from the NASA Astrophysics Data System (ADS) database\footnote{\url{https://ui.adsabs.harvard.edu/}}. Astro-mT5 \citep{ghosh-etal-2022-astro} identifies 32 named entities from the astrophysics literature data provided by the DEAL SharedTask team. \cite{sotnikov2023language} extracts named entities from the Gamma-ray Coordinates Network (GCN) circulars, including event ID, object name, observed event, observed object, and physical phenomena.  AstroLLaMA fine-tunes LlaMA-2 \citep{nguyen2023astrollama}, leveraging a corpus of 300,000 astronomy abstracts. Downstream tasks include text generation (e.g., completing abstracts) and embedding space quality (e.g., where similar abstracts can be retrieved by computing the cosine similarity between the vector embeddings). 

AstroLLaMA-chat is an enhanced version of AstroLLaMA, focused on the task of question answering \citep{perkowski2024astrollama}. Pathfinder \citep{iyer2024pathfindersemanticframeworkliterature} is a machine-learning framework for literature review and knowledge discovery in astronomy, focused on semantic searching with natural language. It is an open-source tool that uses a corpus of more than 300,000 abstracts from NASA ADS. \cite{ciucua2023harnessing} focused on hypothesis generation about Galactic Astronomy using a primal LLM for in-context learning with 1000 papers from NASA ADS, and an adversarial LLM model to critique the idea. 

Text summarization is the task of generating a shorter version of a document that preserves the important information and key points while reducing the length. \cite{platform_astronomy} developed Skyportal, an open-source package designed to efficiently discover interesting transients, including multimessenger features, manage follow-up, perform characterization, and visualize results. Skyportal uses ChatGPT to provide human-readable summaries of individual sources, using redshift, classifications, and comments. 

Recently, LLMs have been tested as AI agents used as research assistants. ReplicationBench \citep{ye2025replicationbench} is a benchmark for evaluating whether AI agents can replicate entire research papers in astronomy. This includes the experimental setup, derivations, data analysis, and codebase. This is a challenging task, and the best LLM scored under 20\%. ASTROVISBENCH \citep{joseph2026astrovisbench} is a benchmark for scientific computing and visualization in astronomy. An evaluation of state-of-the-art LLMs revealed a significant gap in their ability to serve as effective research assistants.
   
Another relevant application of LLMs is the task of text-to-SQL (T2S) parsing, which aims to convert a natural language (NL) question about a database to its corresponding structured query language (SQL) query, that must be executable. ScienceBenchmark \citep{10.14778/3636218.3636225} is a database containing NL questions and SQL queries for benchmark purposes, which includes partially the Sloan Digital Sky Survey (SDSS) database\footnote{\url{https://skyserver.sdss.org/}}. However, the highest execution accuracy obtained in SDSS using LLMs reached only 33\%. The authors conclude that the proposed benchmark is highly challenging and that the T2S task is far from being solved, especially for complex, scientific datasets.

In this work, we develop a T2S system based on LLMs to be applied to the database of the Automatic Learning for the Rapid Classification of Events (ALeRCE). ALeRCE \citep{forster2021automatic, carrasco2021alert, sanchez2021alert} is a Chilean-led astronomy broker that is processing the alert stream from the Zwicky Transient Facility \citep[ZTF;][]{bellm2018zwicky}, and it has been selected as one of the Community brokers for the Vera C. Rubin Observatory and its Legacy Survey of Space and Time \citep[LSST;][]{ivezic2019lsst}. Currently, ALeRCE has more than 27,000 users from 139 countries\footnote{Estimation from Google Analytics}. The ALeRCE database\footnote{\url{https://science.alerce.online/}} can be accessed through the ALeRCE ZTF explorer\footnote{\url{https://alerce.online/}}, the SN hunter\footnote{\url{https://snhunter.alerce.online/}}, a ZTF API with simplified queries\footnote{\url{https://api.alerce.online/ztf/v1}}, or directly via PostgreSQL.

However, producing SQL queries requires computational expertise and time to learn a new database. The Astronomical Data Query Language (ADQL) \footnote{\url{https://www.ivoa.net/documents/ADQL/20180112/}} includes astronomy-specific geometry functions, but its use has remained highly technical. Under the idea of democratizing access to data in the Rubin era, we aim to develop a system where the input is a question in NL text for the ALeRCE database, and the output is the corresponding executable SQL query. 
To this end, we propose a step-by-step framework for performing the T2S parsing task using in-context learning with LLMs, and compare its performance with a direct inference system. In Sect. \ref{sec:background}, we present a historical context of T2S parsing and related work. In Sect. \ref{sec:methodology}, we describe our methodology, wherein we construct a dataset of NL questions and their corresponding SQL queries for the ALeRCE database, which we make publicly available. Section \ref{sec:results} presents our results, where we evaluate the performance of thirteen LLMs using in-context learning and analyze the errors. In Sect \ref{sec:discussion}, we discuss possible improvements such as function calling and structured outputs. Finally, we draw conclusions in Sect. \ref{sec:conclusions}.

\section{Background}
\label{sec:background}

\subsection{Background on text-to-SQL parsing}

The T2S task can be defined as follows \citep{qin2022survey_txt2sql}: Given a NL question Q and the corresponding database schema S=(T,C), where T stands for tables and C for columns, the goal is to generate an SQL query Y, that when executed will return results that match the user's intent \citep{katsogiannis2023survey}. The question Q is a sequence of |Q| tokens. The database consists of |T| tables and |C| columns. Each table is described by its name, which contains multiple words. Each column within a table is also described by words. Schema linking, which consists of aligning entities in questions to tables or columns is the most important topic of T2S problems \citep{bird}. In addition, external knowledge can be used to improve the model's comprehension of the problem. Typically, this refers to domain-specific knowledge, e.g., astronomy, as well as mathematical computation. 
The T2S task presents two kinds of challenges. The first is related to understanding the NL question. The NL is inherently ambiguous \citep{katsogiannis2023survey}, i.e., the formulation of expressions is open to more than one interpretation. The second is that SQL has a strict syntax, which leads to limited expressivity compared to NL. This makes building the syntactically and semantically correct SQL query difficult based on the underlying database schema.

For evaluating T2S systems, several database benchmarks have been created, such as Spider \citep{spider}, BIRD \citep{bird}, and ScienceBenchmark \citep{10.14778/3636218.3636225}. The Spider dataset contains 10,181 NL questions and 5,693 unique SQL queries over 200 databases belonging to 139 different domains, allowing it to deal with the cross-domain T2S parsing task \citep{qin2022survey_txt2sql}. The queries are divided into four levels of difficulty: easy, medium, hard, and extra hard. However, Spider mainly contains simple databases with few tables, columns, and entries, which rarely reflect real-world problems \citep{10.14778/3636218.3636225}. BIRD is a large-scale cross-domain benchmark covering 95 databases from 37 domains. \cite{bird} performed an error analysis using GPT-3.5 Turbo in the T2S task on BIRD. The authors found that the most common error is wrong schema linking (41.6\%), i.e., erroneous association of tables and columns to the NL question. The second most common error was misunderstanding database content (40.8\%), i.e., failing to recall the correct database structure (e.g., using a column that does not belong to a specific table) or generating fake schema items (e.g., calling a table or column that does not belong to the database). However, the previous two database benchmarks do not contain domain-specific knowledge of astronomy. ScienceBenchmark is a new database benchmark that contains three real-world and domain-specific databases: research policy-making, astrophysics, and cancer research. A subset of the SDSS database is used (5 out of 10 tables) due to limitations in the number of tokens allowed by LLMs. For this database, 200 NL/SQL pairs were manually generated by a team of domain and SQL experts, and data augmentation was carried out. The authors found that state-of-the-art T2S algorithms that obtained over 80\% of execution accuracy on Spider, achieved only 21\% execution accuracy on SDSS. In addition, GPT-3.5 with prompting achieved only 33\% execution accuracy. This poor performance is presumably because SDSS is a domain-specific database that includes many numerical values and requires the use of functions and mathematical operators. In addition, it contains many column names and values labeled with domain-specific abbreviations, e.g., z for redshift. The authors concluded that the current state-of-the-art approaches do not perform well on real-world problems, and that we need domain-specific benchmarks for training and evaluating T2S systems in scientific domains.

\subsection{Prompt engineering for the T2S task }

\begin{figure}[t]
   \centering
   \includegraphics[width=0.6\linewidth]{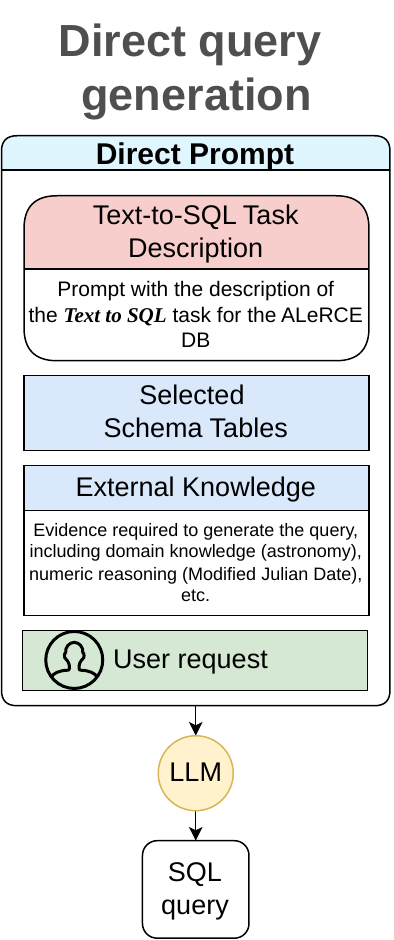}
   \caption{Four components of a basic prompt for the T2S task. This approach is called direct query generation.}
   \label{fig:incont_learning_schema}
\end{figure}

LLMs are currently language models with billions of parameters, trained on massive amounts of text data. According to \cite{zhao2023surveyllm}, LLMs show emergent abilities such as in-context learning, instruction following, and step-by-step reasoning. In-context learning allows LLMs to convert a NL question into a SQL query using a prompt text, i.e., it is an inference task (there is no training or fine-tuning). The prompt usually includes four components:  a task instruction, a NL question, the database schema, and external knowledge \citep{chang2023prompt}. Figure~\ref{fig:incont_learning_schema} shows the four components of a basic prompt for the T2S task, which is the basic inference task using in-context learning, and serves as a baseline for comparison purposes. We call this scheme Direct Query Generation.

Several studies have shown that the ability of LLMs for complex reasoning can be improved using step-by-step reasoning \citep{kojima2022large}. There are several prompting techniques for multi-step reasoning, such as Chain of Thought  \citep[CoT;][]{NEURIPS2022_9d560961}, which is able to guide LLMs through a logical reasoning chain, mimicking how humans break down problems into logical intermediate steps \citep{sahoo2024systematic}. A variant is the Logical Chain-of-Thought \citep{liu2023logicot} prompting, which includes effective verification mechanisms to reduce logical errors and hallucinations through a think-verify-revise loop \citep{zhao2024enhancing}. Regarding code generation, Structured Chain-of-Thought (SCoT) prompting incorporates program structures (sequence, branch, and loop structures) into reasoning steps \citep{li2025structured}.

The DIN (Decomposed In-Context)-SQL method \citep{NEURIPS2023_72223cc6} decomposes the text-to-SQL task into four modules: 1) schema linking, 2) query classification and decomposition, 3) SQL generation, and 4) self-correction. A prompt-based module is designed for schema linking, which includes ten random samples from the training set of the Spider dataset. For each mention of a column name in the question, the corresponding column and its table are selected from the given database schema. The second module classifies each query into one of three classes: easy, non-nested complex, and nested complex. The class labels are essential for the query generation module, which uses different prompts for each query class. A simple few-shot prompting with no intermediate steps is performed for questions in the easy class. In addition to class labels, the module also detects the tables to be joined for non-nested and nested queries. The nested complex class is the hardest and requires several intermediate steps before generating the final answer. The prompt for this class is designed in a way that the LLM first solves sub-queries and then uses them to generate the final answer. Finally, a self-correction module is added, where only buggy code is provided to the model, and it is asked to fix the bugs in a zero-shot setting.

The DAIL-SQL \citep{10.14778/3641204.3641221} proposes a new prompt engineering method focused on question representation and in-context learning for text-to-SQL. Among the question representations studied are a basic prompt (table schema and question in NL), text-representation prompt (add task instruction to the basic prompt), OpenAI Demonstration Prompt (ODp) (the instruction is more specific, constraining how the model must answer), and Code Representation Prompt (CRp) (it presents the T2S task in SQL syntax, e.g., the NL questions appear inside SQL comments). Regarding in-context learning, DAIL-SQL focuses on example selection and example organization. Among the example selection strategies are random selection, question-similarity selection (select k examples with the most similar questions), query-similarity selection (select k examples similar to the target SQL query), and DAIL Selection (consider both questions and queries to select examples). 
Experiments were run with the Spider and Spider-Realistic datasets. In the zero-shot scenario, the ODp representation performed best in GPT-3.5 Turbo, but the basic prompt performed best in GPT-4. Adding foreign key information, i.e., the relationships between tables in a relational database, significantly improves the execution accuracy of LLMs. Adding the rule to generate SQL queries “with no explanation” consistently improved the performance of all LLMs. DAIL selection generally outperformed other example selection strategies. 
Similar approaches to DIN-SQL and DAIL-SQL have been presented in MAC-SQL \citep{wang2025mac}, and C3 \citep{dong2023c3}.

\section{Methodology}
\label{sec:methodology}

\subsection{ALeRCE database}

As of early 2026, the ALeRCE database has 25 tables and 304 columns; see the entity-relationship diagram in Appendix \ref{appendix:alerce_db}. It contains information about astrophysical objects, including global and band-dependent statistics, time- and band-dependent flux evolution, cross-matches, and machine-learning probabilities. For example, the ALeRCE light curve classifier \citep{sanchez2021alert} computes machine learning probabilities for 15 classes: SNIa, SNIbc, SNII, SLSN, QSO, AGN, Blazar, CV/Nova, YSO, LPV, EB, DSCT, RRL, CEP, and Periodic-Other.
The main tables are ‘object’ (22 columns), ‘probability’ (6 columns), ‘magstat’ (28 columns), ‘detection’ (30 columns), ‘non\_detection’ (4 columns), 'forced\_photometry' (42 columns), and ‘feature’ (5 columns). These tables contain information about each object, including time and spectral band statistics. The organization of the database is centered on the ‘object' and ‘probability’ tables, which contain the attributes and classification of the objects, respectively, and which are requested in most of the queries requiring JOIN\footnote{Joining in SQL means retrieving data from two or more tables based on a common field.} commands or sub-queries.  
To collect NL/SQL pairs, we first included 21 examples generated by the ALeRCE team and presented in workshops with the aim of teaching users how to interact with the database. In addition, a group of 10 astronomers were tasked with generating real-world questions, with the only restriction being that the information to answer these questions must be obtainable from the ALeRCE database alone. In this way, another 89 examples were generated. Finally, an SQL expert built the target gold queries, resulting in a total of 110 NL/SQL pairs obtained. Listing \ref{lst:example_query} shows an example of a query, where ranking=1 means to take the highest probability only.

\begin{lstlisting}[language=SQL, caption={Example of a question in natural language and its corresponding SQL query.}, label={lst:example_query}]
<@\textcolor{red}{User Request}@>: 
<@\textcolor{black}{Give me the objects classified as YSO by their}@>
<@\textcolor{black}{lightcurves (with a probability higher than 0.7)}@>
<@\textcolor{green}{Query}@>: 
SELECT
    oid, probability
FROM
    probability
WHERE
    classifier_name='lc_classifier' AND
    class_name='YSO' AND
    ranking=1 AND
    probability > 0.7
\end{lstlisting}

Most of the SQL queries included in our gold set represent typical requests found in the literature, which show how the community has already used ALeRCE. These include: requesting light curve detections for given objects \citep[e.g.,][]{magee2023search, gomez2023search, muller2025analyzing, alexander2026multiwavelength}; radial queries \citep[e.g.,][]{pomeroy2024search}; crossmatches with objects from a given class in one of the ALeRCE classifiers \citep[e.g.,][]{lopez2022confirming, arevalo2024newborn, oshikiri2024search}. Meanwhile, other queries are highly specific, including e.g., queries to individual features, or filters through tables other than object and probability (data quality, the Panoramic Survey Telescope \& Rapid Response System (PS1)\footnote{\url{https://catalogs.mast.stsci.edu/panstarrs}} data, Wide-field Infrared Survey Explorer (WISE)\footnote{\url{ https://irsa.ipac.caltech.edu/Missions/wise.html}} data, and Solar System data); while these requests are expected to be less frequent, they allow us to clean object samples and/or get a broader understanding of the selected objects, and thus deserve to be included in our gold set.

\subsection{Query classification}

Queries were classified into three difficulty levels: simple, medium, and hard. This was done first to evaluate how the LLM performance varies with query difficulty and, second, to allow generating a different prompt depending on the level of difficulty, as explained in Sect. \ref{framework}. To this end, we define the tables ‘object’, ‘probability’, and ‘magstat’ as basic tables (the most commonly used) and the remaining tables as general. Definitions of the categories are as follows:
\begin{itemize}
  \item Simple: queries using up to two basic tables, or queries using a single general table.
  \item Medium: queries that require either one general and one basic table, two general tables, or three basic tables. This category also includes queries that contain two simple sub-queries—each comparable in complexity to a simple query—or one sub-query of higher complexity.
  \item Hard: Any query requiring more than three tables or using more than one non-simple sub-query for optimization.
\end{itemize}

Based on these, the ALeRCE dataset of 110 NL/SQL pairs was partitioned into a development set of 58 examples and a testing set of 52 examples. Note that when using in-context learning, there is no need for a training set, since there is no parameter adjustment. The development set is used for developing and evaluating diverse prompt methods. The testing set is exclusively used for the final evaluation.
The 58 examples of the development set consisted of 32 simple, 14 medium, and 12 hard queries. The 52 examples of the testing set were composed of 32 simple, 10 medium, and 10 hard queries. Examples of simple, medium and hard queries are given in Appendix \ref{appendix:examples_diff}.

\subsection{LLM models}
Table \ref{tab:used_models} describes the thirteen LLM models used in this work. This includes the full name of the API, the abbreviations of names used in what follows, and the context length. The latter is the maximum number of tokens (input + output) that an LLM can process in a single forward pass.

\begin{table}[b]
    \caption{Description of the LLM models.}
    \label{tab:used_models}
    \footnotesize
    \begin{tabular}{l l l}
    \hline \hline
    API Model & Abbreviation & \begin{minipage}{1.8cm}
        \ \\ Context \\ length \\ \
        \end{minipage}\\ \hline
    gpt-4o-2024-11-20 & GPT-4o & 128\,000 \\ 
    gpt-4.1-2025-04-14 & GPT-4.1 & 1\,048\,576 \\ gpt-5-2025-08-07 & GPT-5 & 400\,000 \\ 
    gpt-5.2-2025-12-11 & GPT-5.2 & 400\,000 \\ 
    gpt-5.2-codex & GPT-5.2-Codex & 400\,000 \\ 
    gpt-5.3-codex & GPT-5.3-Codex & 400\,000 \\ 
    claude-3-7-sonnet-20250219 & Claude 3.7 & 200\,000 \\ 
    claude-sonnet-4-5-20250929 & Claude Sonnet 4.5 & 200\,000 \\ 
    claude-opus-4-6 & Claude Opus 4.6 & 200\,000 \\ 
    gemini-2.5 flash & Gemini 2.5 Flash & 1\,048\,576 \\ 
    gemini-2.5 pro & Gemini 2.5 Pro & 1\,048\,576 \\ 
    gemini-3-flash-preview & Gemini 3 Flash & 1\,048\,576 \\ 
    gemini-3.1-flash-lite-preview & Gemini 3.1 Flash & 1\,048\,576 \\ \hline
    \end{tabular}
   
\end{table}

\subsection{Basic prompt}

The basic inference model shown in Fig. \ref{fig:incont_learning_schema} is used as a baseline for comparison purposes. In the zero-shot case, the information given to the model includes the requested T2S task, the database schema, external knowledge (which is query-dependent), and general information about the database structure. The format of the zero-shot prompt is shown in Listing \ref{lst:zero_shot_prompt}. Optionally, a few examples could be added to the prompt, as in the so-called few-shot case. In this work, we mainly deal with the zero-shot case. The few-shot case is presented in Sect. \ref{sec:discussion} for the best LLM models.

\begin{lstlisting}[language=Python, caption={Zero-shot prompt format.}, label={lst:zero_shot_prompt}]
{General_Task}

# Context:
{General_Context}

# The Database has the following tables that can be used to generate the SQL query:
{tables_schema}

{Final_Instructions}

# Important Information for the query
{External_Knowledge}
# User Request: ''{query}''
\end{lstlisting}

\subsection{Proposed framework for the T2S task} \label{framework}
\begin{figure}
   \resizebox{\hsize}{!}{\includegraphics{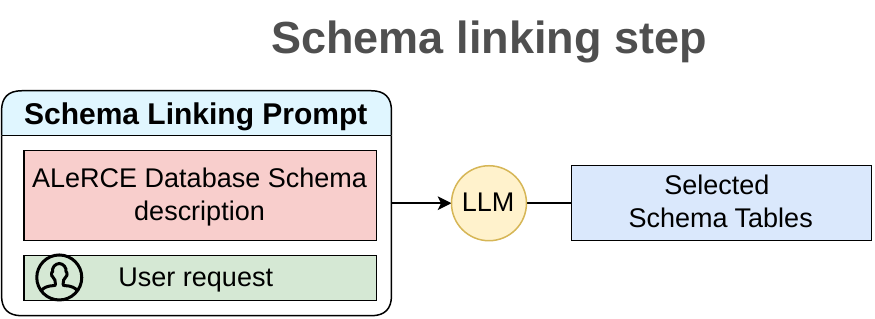}}
   \caption{Diagram of the schema linking, the first step of the proposed framework for the T2S task.}
   \label{fig:framework_schema_linking}
   \end{figure}

\begin{figure}
   \resizebox{\hsize}{!}{\includegraphics{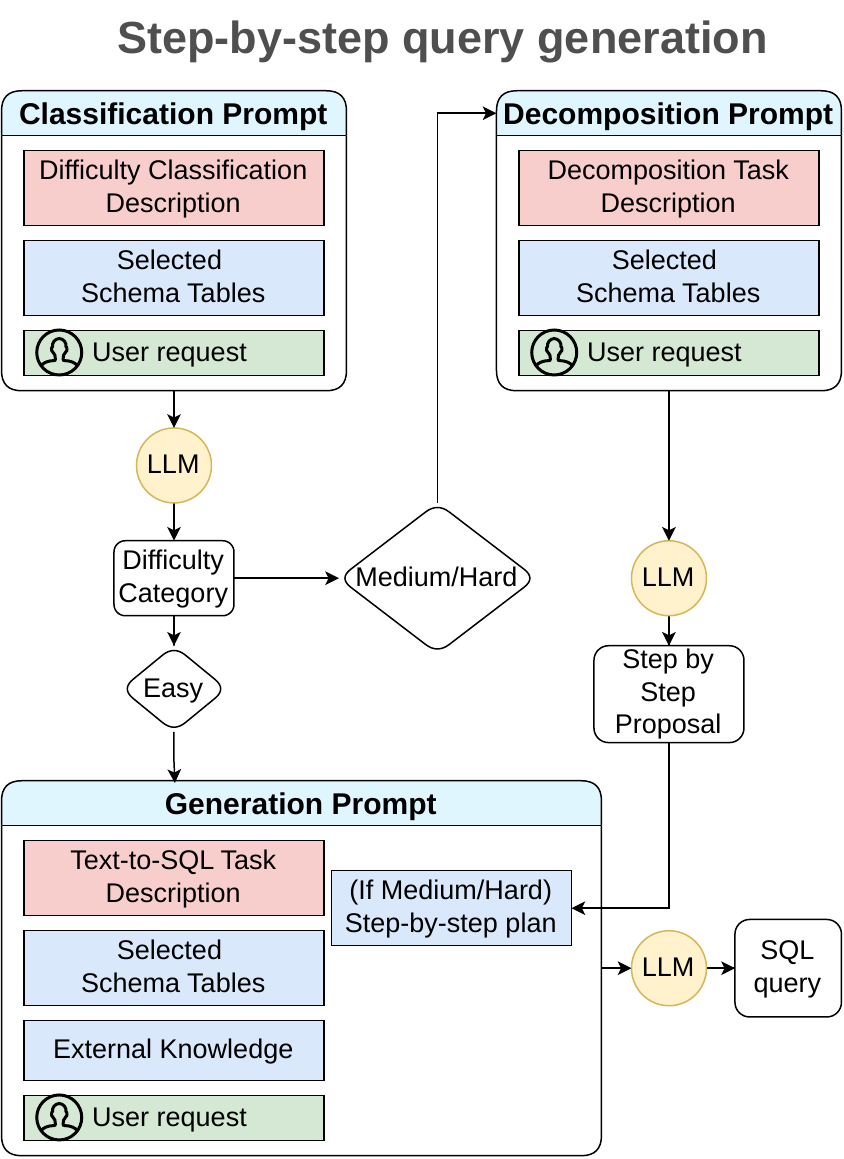}}
   \caption{Step-by-step query generation approach includes a classification prompt and a decomposition prompt. The classification prompt aims to discriminate between simple, medium, and hard queries. The decomposition prompt divides the problem into sub-problems for medium and hard queries. Simple queries skip the decomposition and go directly to the generation prompt.}
   \label{fig:framework_schema}
   \end{figure}

\begin{figure}
   \resizebox{\hsize}{!}{\includegraphics{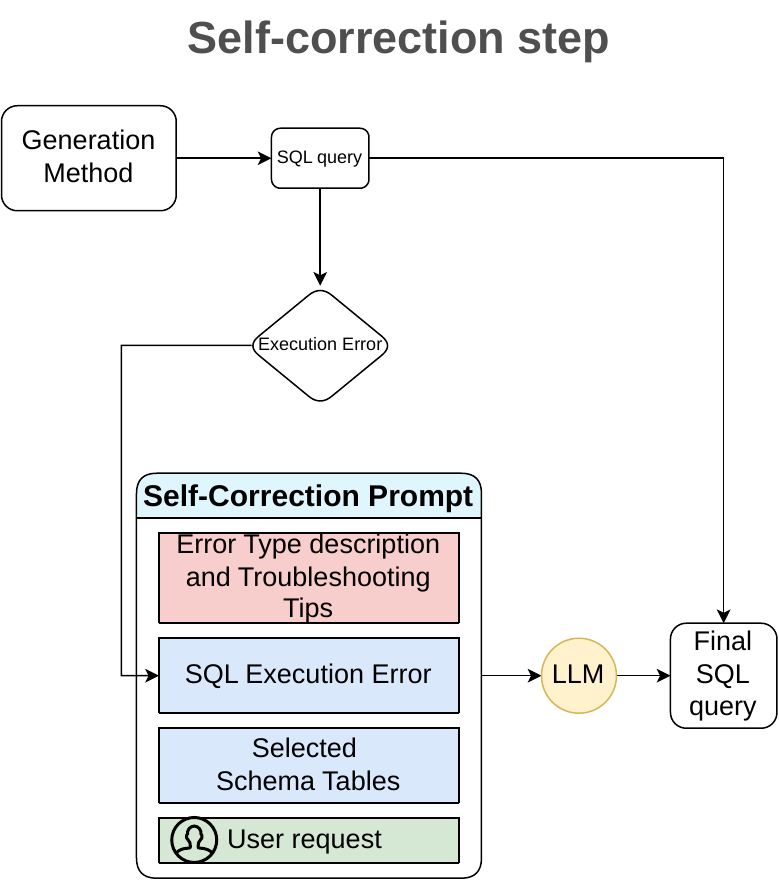}}
   \caption{Self-correction step generates a prompt specific to each type of SQL error: time-out, non-existing structures, or general errors.}
   \label{fig:framework_selfcorrection}
   \end{figure}

Our end goal is to develop a practical virtual assistant for ALeRCE users.  With this aim, we decompose the prompt into different steps that run autonomously. Our approach is inspired by previous work such as DIN-SQL \citep{NEURIPS2023_72223cc6} and DAIL-SQL \citep{10.14778/3641204.3641221}.
The proposed framework for performing the T2S task in the ALeRCE database has four steps: schema linking, classification, decomposition, and self-correction. The first step, schema linking, is shown in Fig.~\ref{fig:framework_schema_linking}. Steps 2 and 3 are shown in Fig.~\ref{fig:framework_schema}. Finally, Fig.~\ref{fig:framework_selfcorrection} shows the self-correction step, the last step of the proposed framework. 
A detailed description of the four steps of the proposed framework is as follows:
\begin{enumerate}
       \item Schema Linking: The aim is to align the question with the database schema, i.e., identify the required tables and columns. With this aim, an independent module was designed to extract the required structure from the database; see Fig.~\ref{fig:framework_schema_linking}. This is carried out in two stages. The first identifies the required tables, and the second determines the columns. Each stage is performed using zero-shot prompting. For the task of extracting the tables, the model is provided with a question, a description of the tables in the database, and it is instructed to perform the schema linking task. The columns are extracted independently for each of the previously selected tables.  A different API call is started for each table, with a prompt describing the columns contained in the table. The outputs of this step are the selected schema tables. The prompt for the schema linking is shown in the Appendix \ref{appendix:self_corr_prompts}, Listing \ref{lst:schema_linking_prompt}.
    
    \item Classification: This step determines the level of difficulty of the query.  The prompt for classifying the level of difficulty uses the structure shown in the Appendix \ref{appendix:self_corr_prompts}, Listing \ref{lst:diff_Class_prompt}. The prompt includes the definition of the query classes by difficulty, the tables and columns obtained in the schema linking stage, and the user query. 
    \item Decomposition: As shown in Fig.~\ref{fig:framework_schema}, a decomposition prompt is performed for the medium and hard queries. For simple queries, decomposition is not needed; they go directly to the generation prompt block. 
    Otherwise, we decompose the problem into sub-problems. This is useful for complex queries that require different sub-queries.   We tested least-to-most prompting \citep{zhou2022least} and decomposed prompting \citep{khot2022decomposed}. A new module is used to decompose the queries that require complex sub-queries or JOINs. The input to this module is the information extracted in the previous stages. 
    \item Self-Correction: Recent LLM models, such as GPT-5.2-Codex, Claude Opus 4.6 or Gemini 3 Flash, are able to correct or improve their original response. A new agent can be used to get feedback about previous responses. The self-correction module addresses the three main types of execution errors found in the generation of queries: timeout; non-existing structures, such as confusion of column and table names, or non-existent columns, tables, functions, etc.; and general errors such as syntax, missing FROM-clauses, and so on. A schema of the self-correction step is shown in Fig.~\ref{fig:framework_selfcorrection}.
    The self-correction step consists of a single iteration and is triggered only when the generated SQL query produces an execution error, so it is not applied to every query. This step is particularly effective at fixing syntax errors and undefined or ambiguous references. Timeout errors can also be corrected when the model identifies the minimal conditions and clauses needed to produce a valid result within the time limit. A different zero-shot prompting is used for each type of error. For the timeout error (more than 2 minutes), it is assumed that there is an optimization problem. The agent is asked to improve or re-order the query according to the error found. In addition, advice on avoiding this type of error is provided. The main structure of the prompt for timeout errors is shown in Appendix \ref{appendix:self_corr_prompts}, Listing \ref{lst:timeout_selfcorr_prompt}.
A similar structure is used for general errors and non-existent structures, but focusing the prompt on the syntax and the database structure errors.

\end{enumerate}

\subsection{Prompt engineering}

The prompt is separated into system messages and user messages, as done in OpenAI\footnote{\url{https://platform.openai.com/docs/guides/prompt-engineering/\#messages-and-roles}}. The system messages include the general task, the ALeRCE context, and the database schema. The user messages include the user's question and the external knowledge, if needed. For the general task, a personality as a data scientist expert is given:

\begin{quote}
``As an SQL expert with a willingness to assist users…''
\end{quote}

The general context describes the ALeRCE pipeline, the objects and classes used, as well as details on how ALeRCE works (including the probability assigned to objects). The following is an example:

\begin{quote}
``ALeRCE Pipeline Details \\ - Stamp Classifier (denoted as `stamp\_classifier'): All alerts related to new objects undergo stamp-based classification. \\ - Light Curve Classifier (denoted as `lc\_classifier'): A balanced hierarchical random forest classifier employing four models and 15 classes. \\
...''
\end{quote}

For the database schema, commands to create the ALeRCE SQL tables are given: 

\begin{quote}
``CREATE TABLE object (/* this is the most important table. It contains the main statistics of an object, independent of time and band */ \\
oid VARCHAR PRIMARY KEY \\
...)''
\end{quote}

In addition, instructions on the format of the answers are given:

\begin{quote}
``Answer ONLY with the SQL query ...'', ``... Do NOT CHANGE the names of the tables or columns unless the user explicitly asks you to do ...''
\end{quote}

For the step-by-step method, different prompts were generated for each step. For the schema linking step, the prompt included a brief description of the task, a description of the tables of the database, and specifications regarding when to select the object, probability, and feature tables. This is because the object and probability tables are designed for frequent use and have been indexed accordingly, while the features table is designed for more advanced use cases and is generally more expensive to query.
For the classification step, a prompt with the classification task, a description of the three query categories, the table schema selected from the schema linking step (including column descriptions), instructions associated with the classification, and the user request are used.

For the decomposition step, the query generation was split into two stages. In the first stage, the LLM is asked to generate a step-by-step plan to generate the query, including all requirements, conditions, and details.
In the second stage, the LLM is requested to generate the query following the plan generated in the first stage.
For medium queries, a prompt example for the first stage is the following:

\begin{quote}
``Your task is to DECOMPOSE the user request into a series of steps required to generate a PostgreSQL query ...''.
\end{quote}

For hard queries, a higher emphasis is put on the details and order of the query structure, as follows:

\begin{quote}
``The request is a very difficult and advanced query, so you will need to use JOIN, INTERSECT, and UNION statements, together with Nested queries. It is very important that you give every possible detail in each step, ...''.
\end{quote}

Following BIRD \citep{bird}, we add external knowledge to the prompts in order to provide the information needed to solve the problem, but that is not present in the dataset, such as numeric reasoning knowledge, domain knowledge, and synonym knowledge. This external knowledge is specific to each query. The following is an example of the external knowledge required to solve a simple query:
\begin{quote}
Request: ``Query all objects that were first classified as SN by the stamp classifier between August 17 and August 21, 2024, with a probability greater than 0.5 or at least two detections.''\\
External Knowledge:\\
-- MJD date for August 17 2024 = 60173.0\\
-- MJD date for August 21 2024 = 60177.0
\end{quote}

Notice that we used the same prompts for all LLMs, following established prompt
engineering practices. This independence of the LLM provider allows us to make fair comparisons and to use these prompts with new versions of the LLMs, when the old models become deprecated. In addition, it allows the users to select an LLM of their own choice.

\subsection{Evaluation metrics }

The most common metric used in T2S is the execution accuracy (EX). EX is the proportion of questions in the test set for which the execution results of both the predicted and ground-truth inquiries are identical \citep{bird}. In this work, we propose a modified EX, to consider partially correct results. We compute the set of row identifiers, called ids, in the ground truth, $O_t$, as well as the set of the predicted ids, $O_p$.  In the ALeRCE database, examples of row identifiers are `object identifier' (oid), candidate identifier (candid), oid\_catalog, objectidps1 (object id closest source from PS1 catalog), classifier\_name, or count.  

For this binary classification problem, precision and recall are calculated. For each predicted query ($k$, $l$), where $k$ is the user query identifier and $l$ is the query run identifier (e.g., an index from 0 to 9 for 10 runs of the user query $k$), the recall metric is calculated by counting the predicted identifiers ($id_i^{k, l}$, where $i$ is the row number among the predicted values) that match the set of true ids ($O^{k}_t$), divided by the cardinality of the set of true ids, $|O^{k}_t|$:
\begin{equation}
    r_{k,l} = \frac{\sum_{i \in \text{pred}} \mathit{score}(id_i^{k, l}, O^{k}_t)}{|O^{k}_t|}\label{eq:EPi},
\end{equation}
where
\begin{equation}
    \mathit{score}(id,O) =  \begin{cases}
       1 ,& id \in O \\
       0 ,& id \notin O
    \end{cases}  \label{eq:scoreEP}.
\end{equation}

Likewise, for each user query $k$ and query run $l$, the precision metric is calculated by counting the matches between the true ids ($id_i^{k, l}$, where $i$ is the row number among the true values) and the set of predicted ids ($O^{k, l}_p$), and dividing this amount by the cardinality of the set of predicted ids, $|O^{k,l}_p|$:
\begin{equation}
    p_{k,l} = \frac{\sum_{i \in \text{true}} \textit{score}({id}^{k, l}_i,O^{k,l}_p)}{|O^{k,l}_p|}\label{eq:ERi}.
\end{equation}

The same can be done in the column dimension, where, for example, an oid can have associated values such as firstmjd, lastmjd, ndet (number of detections), and so on. Thus, we define $r_{k, l}^{rows}$ and $p_{k, l}^{rows}$ as the recall and precision computed along the row identifiers. Likewise $r_{k, l}^{cols}$ and $p_{k, l}^{cols}$ are the recall and precision computed along the column identifiers.

In this work, we assume that an answer containing all and only the expected rows is correct. On the other hand, we assume that an answer containing all but not necessarily only the expected columns is correct. This is because the latter will not significantly impact the user experience. According to this assumption, we evaluate all prompting methods using a metric of perfect matching rates ($PM$), defined as follows:
\begin{equation}
    PM = \frac{1}{n_{\rm queries} ~ n_{\rm runs}} \sum_{k \in {\rm queries}} \sum_{l \in {\rm runs}} N_{\rm perfect}(k, l),
\end{equation}
where $n_{\rm queries}$ is the total number of queries, and $n_{\rm runs}$ is the number of runs per query, typically 10. $N_{\rm perfect}(k, l)$ is defined as:
\begin{equation}
    N^{\rm rows}_{\rm perfect} (k,l) = \begin{cases}
        1, \text{ if $p^{\rm rows}_{k,l}$ = $r^{\rm rows}_{k,l}$ = 1} \\
        0, \text{ otherwise}
    \end{cases},
    \label{eq:nrow}
\end{equation}
\begin{equation}
    N^{\rm cols}_{\rm perfect} (k,l) = \begin{cases}
        1, \text{ if $r^{\rm cols}_{k,l}$ = 1} \\
        0, \text{ otherwise}
    \end{cases},
    \label{eq:ncol}
\end{equation}
for the case of rows and columns, respectively.
Accordingly, we define $PM_{\rm rows}$ and $PM_{\rm cols}$ as metrics for perfect matches among the rows or columns, respectively.

\section{Results}
\label{sec:results}

In this section, we present results comparing the performance of direct vs. step-by-step (sbs) query generation methods, with and without self-correction, as a function of query difficulty across the thirteen LLMs. In addition, we analyze the errors obtained.
Table \ref{tab:combined_vertical_results_by_exp} in Appendix \ref{appendix:pm_rates} shows the PM results. For evaluation, we conduct ten runs and compute the average PM across the entire query set. We
report the mean and standard deviation across these runs. Consistency varies across LLM models, even
when using a zero temperature. For earlier models, such as Claude 3.7, GPT-4o and GPT-4.1, SQL
outputs vary more noticeably for medium and hard queries, where more tokens are required and thus the
probability of divergence increases (see the standard deviation in Table \ref{tab:combined_vertical_results_by_exp}). Newer models have more deterministic behavior, notably Gemini 3 Flash, Claude Sonnet 4.5, and Claude Opus 4.6.
It can be observed in Table \ref{tab:combined_vertical_results_by_exp} that when comparing the performance of each LLM independently, without (w/o) and with (w/) self-correction, in all cases, the self-correction improves or achieves a similar result to that without self-correction.

\begin{figure}[t]
    \resizebox{\hsize}{!}{\includegraphics{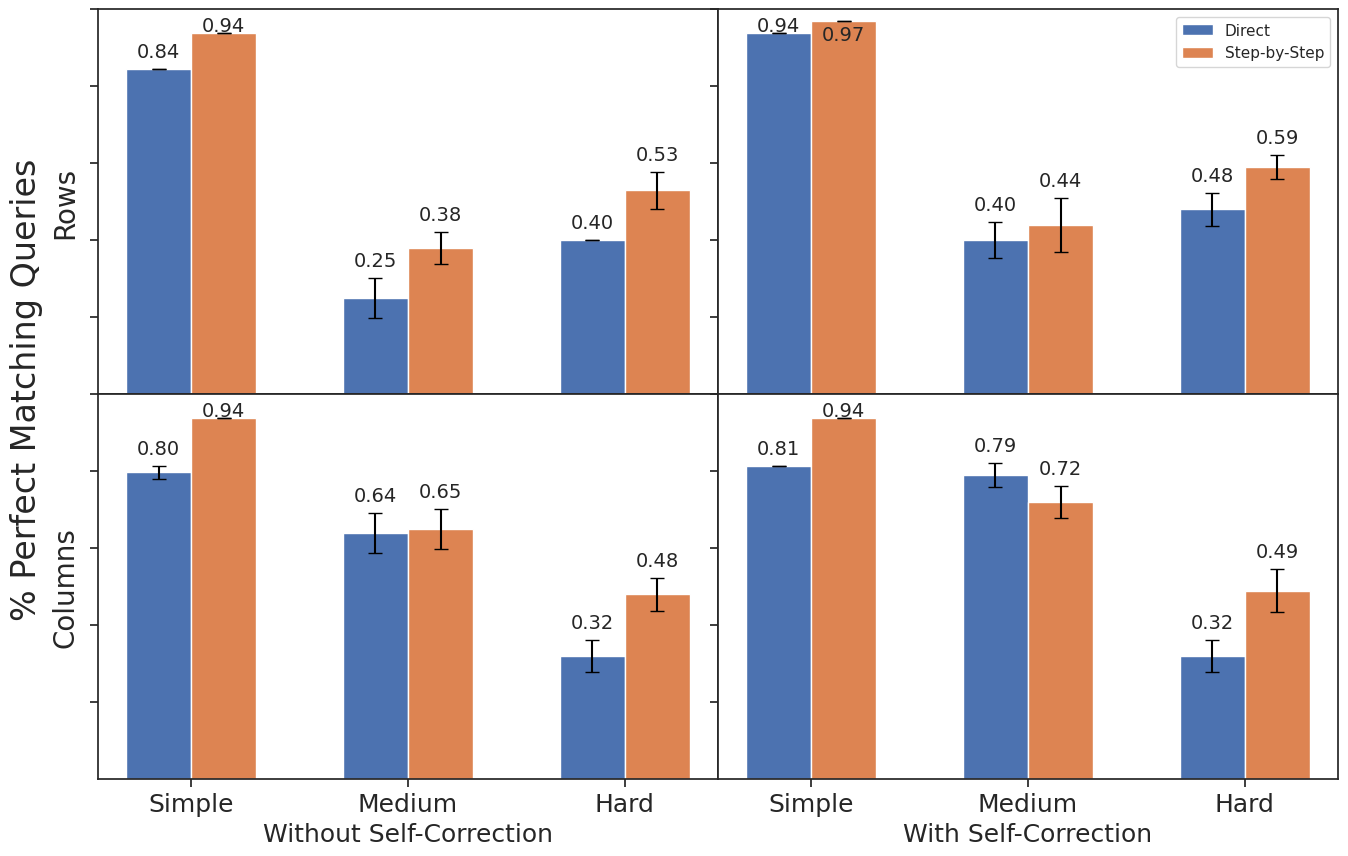}}
    \caption{Claude Opus 4.6 perfect matching performance for rows and columns, with and without self-correction, for the direct and step-by-step query generation methods.}
    \label{fig:nosc_vs_sc_claude46opus}
\end{figure}

Fig. \ref{fig:nosc_vs_sc_claude46opus} shows the performance of Claude Opus 4.6 w/ and w/o self-correction at different levels of query difficulty. When applying self-correction, the results improve or remain the same for all kinds of queries for both the direct query generation and the sbs method. For example, the performance of row ids improves from 0.84 (w/o self-correction) to 0.94 (w/ self-correction) when using the direct query generation method, and from 0.94 to 0.97 when using the sbs query generation method.

Table \ref{tab:general_sig_rank_v2_holm_dense_w_sc} shows a general statistical ranking of the top six LLMs according to their PM performance, using both the sbs and direct query generation methods with self-correction. The top six LLMs in our experiments are: Claude Opus 4.6, Claude Sonnet 4.5, Gemini 2.5 Pro, Gemini 3 Flash, GPT-5.2-Codex and GPT-5.3-Codex. We used an unpaired permutation test to rank the LLMs according to their performance on rows (R) and columns (C), for simple (S), medium (M), and hard (H) queries. A rank of 1 in the table denotes the best-performing LLM. This rank may be assigned to multiple models when pairwise performance differences are not statistically significant. Next, we aggregated the results to obtain a final sum, where a lower number indicates a better ranking. It can be observed that the sbs query generation methods obtained a better ranking than the direct query generation methods in all cases, except for Gemini 3 Flash, which direct query generation version ranked third. Claude Opus 4.6 (sbs) and Gemini 2.5 Pro (sbs) are ranked first and second, respectively.

Table \ref{tab:cost_by_stage} shows the average total cost per run of the whole test set in US dollars, as well as its split into the different stage components, for the top six LLMs. Using the cost of Gemini 2.5 Pro (sbs) as a reference, the cost increases 2.41 times when using Claude Opus 4.6 (sbs), and 1.01 times when using GPT 5.2 Codex (sbs). Gemini 3 Flash has the lowest cost for both the direct and sbs query generation methods. Table \ref{tab:max_tokens_by_stage} in Appendix \ref{additional results} shows the maximum total token consumption, broken down by pipeline stage, LLM model and query generation method. Importantly, all stages remain well within the context window limits of the thirteen evaluated LLMs.

\begin{table*}[htbp!]
\footnotesize
\centering
\caption{General statistical ranking (w/ self-correction).}
\begin{tabular}{l c c c | c c c || c c c}
\toprule
\toprule
Model (Method) & RS & RM & RH & CS & CM & CH & R-SUM & C-SUM & SUM \\
\midrule
Claude Opus 4.6 (SbS) & 1 & 1 & 1 & 1 & 3 & 3 & 3 & 7 & 10 \\
Gemini 2.5 Pro (SbS) & 3 & 1 & 3 & 2 & 2 & 2 & 7 & 6 & 13 \\
Gemini 3 Flash (Direct) & 2 & 2 & 2 & 4 & 2 & 1 & 6 & 7 & 13 \\
GPT-5.2-Codex (SbS) & 3 & 1 & 4 & 2 & 2 & 2 & 8 & 6 & 14 \\
Gemini 3 Flash (SbS) & 2 & 2 & 3 & 2 & 2 & 3 & 7 & 7 & 14 \\
Claude Sonnet 4.5 (SbS) & 2 & 2 & 4 & 3 & 1 & 3 & 8 & 7 & 15 \\
GPT-5.3-Codex (SbS) & 3 & 2 & 3 & 2 & 3 & 2 & 8 & 7 & 15 \\
Claude Opus 4.6 (Direct) & 2 & 2 & 2 & 4 & 2 & 4 & 6 & 10 & 16 \\
Gemini 2.5 Pro (Direct) & 3 & 2 & 3 & 4 & 2 & 3 & 8 & 9 & 17 \\
GPT-5.2-Codex (Direct) & 4 & 2 & 4 & 4 & 2 & 2 & 10 & 8 & 18 \\
Claude Sonnet 4.5 (Direct) & 3 & 1 & 5 & 5 & 3 & 3 & 9 & 11 & 20 \\
GPT-5.3-Codex (Direct) & 5 & 3 & 4 & 5 & 4 & 2 & 12 & 11 & 23 \\
\bottomrule
\end{tabular}
\tablefoot{An unpaired permutation test on per-run perfect-match scores (Holm-Bonferroni correction, $\alpha=0.05$) was performed.} 
\label{tab:general_sig_rank_v2_holm_dense_w_sc}
\end{table*}

\begin{table*}[htbp!]
\footnotesize
\centering
\caption{Average total cost per run in US dollars by LLM and query generation method.}
\setlength{\tabcolsep}{4pt}
\begin{tabular}{lcccccc| c}
\toprule
\toprule
LLM (Query Generation Method) & Total & Schema & Difficulty & Decomp & SQL & Self-Correction & Ratio \\
\midrule
Claude Opus 4.6 (Step-by-Step) & \textbf{2.87} & 0.23 & \textbf{0.43} & \textbf{1.07} & 1.05 & 0.09 & \textbf{2.41} \\
Claude Opus 4.6 (Direct) & 1.55 & 0.23 & --- & --- & \textbf{1.19} & \textbf{0.13} & 1.30 \\
GPT-5.2-Codex (Step-by-Step) & 1.20 & 0.15 & 0.15 & 0.43 & 0.38 & 0.10 & 1.01 \\
Gemini 2.5 Pro (Step-by-Step) & 1.19 & \textbf{0.43} & 0.11 & 0.29 & 0.30 & 0.06 & 1.00 \\
Gemini 2.5 Pro (Direct) & 0.85 & \textbf{0.43} & --- & --- & 0.34 & 0.08 & 0.72 \\
GPT-5.2-Codex (Direct) & 0.82 & 0.15 & --- & --- & 0.56 & 0.11 & 0.69 \\
Claude Sonnet 4.5 (Step-by-Step) & 0.71 & 0.14 & 0.06 & 0.26 & 0.14 & 0.11 & 0.60 \\
GPT-5.3-Codex (Step-by-Step) & 0.61 & 0.07 & 0.07 & 0.19 & 0.21 & 0.06 & 0.51 \\
Claude Sonnet 4.5 (Direct) & 0.39 & 0.14 & --- & --- & 0.16 & 0.08 & 0.33 \\
GPT-5.3-Codex (Direct) & 0.34 & 0.07 & --- & --- & 0.20 & 0.07 & 0.28 \\
Gemini 3 Flash (Direct) & 0.30 & 0.08 & --- & --- & 0.20 & 0.02 & 0.25 \\
Gemini 3 Flash (Step-by-Step) & 0.27 & 0.08 & 0.04 & 0.06 & 0.08 & 0.01 & 0.23 \\
\bottomrule
\end{tabular}
\tablefoot{Values are aggregated over all queries from the test set and averaged over 10 runs. The cost is split into each stage of the query. The ratio compares the total cost against Gemini 2.5 Pro (sbs).}
\label{tab:cost_by_stage}
\end{table*}

\begin{figure}[b]
    \resizebox{\hsize}{!}{\includegraphics{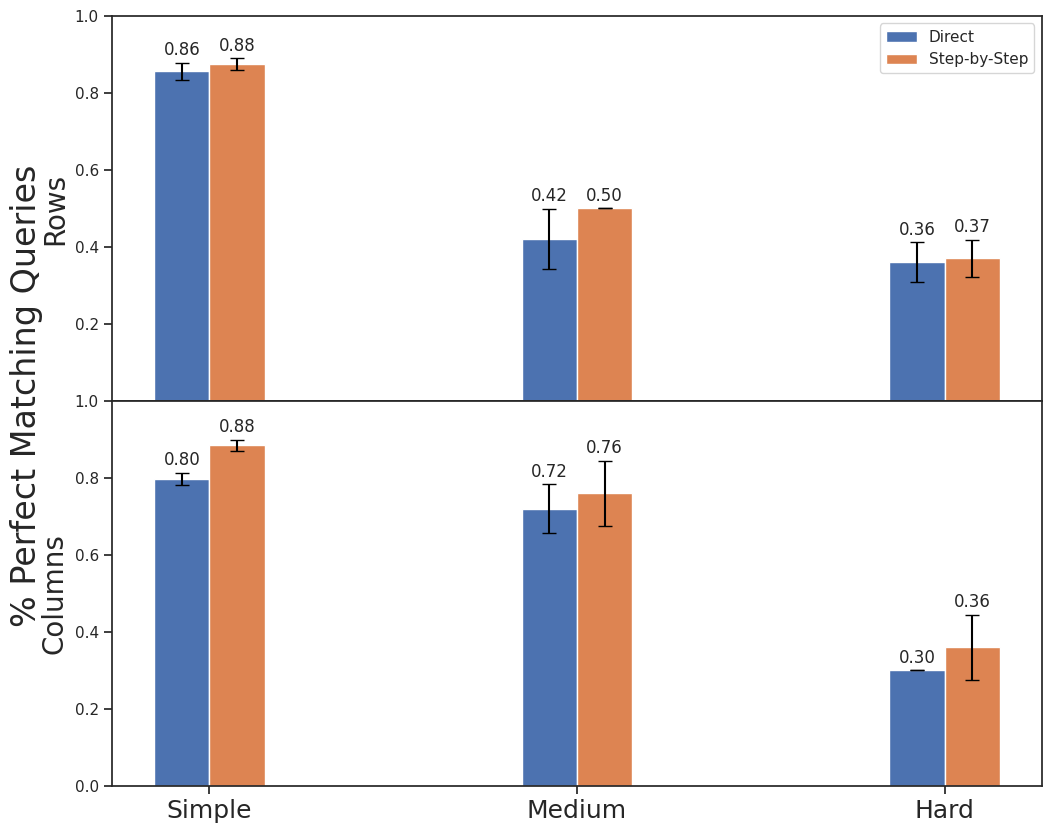}}%
    \caption{PM performance comparison of direct vs step-by-step with self-correction using Gemini 2.5 Pro.}%
    \label{fig:query_gen_comparison5_gemini25pro}%
\end{figure}

Fig. \ref{fig:query_gen_comparison5_gemini25pro} shows the performance of both query generation methods with self-correction using Gemini 2.5 Pro, measured as the rate of PM queries for row and column identifiers, as a function of query difficulty. It can be observed that the proposed sbs framework obtained higher mean values than the direct method in all cases, but there are no statistically significant differences according to the permutation test\footnote{\url{https://rasbt.github.io/mlxtend/user_guide/evaluate/permutation_test/}}, except for medium queries (rows) and simple queries (columns).

The right side of Fig. \ref{fig:nosc_vs_sc_claude46opus} shows the same experiment with Claude Opus 4.6. It can be observed that the proposed sbs framework outperforms the direct method for simple and hard queries. The differences are statistically significant according to the permutation test, except for medium queries where there is a draw. The performance of sbs is high for simple queries (e.g., 0.97 for row ids) and decreases with query difficulty, from 0.44 for medium to 0.59 for hard queries. The columns’ performance follows a similar pattern, highlighting that medium queries achieve a perfect match of 0.72.
Across all queries, the direct query generation method using Claude Opus 4.6 outperformed or performed similarly to the Gemini 2.5 Pro direct query generation method. In addition, when comparing the results of the step-by-step method, Claude Opus 4.6 outperformed or performed similarly to Gemini 2.5 Pro for simple and hard queries.

\begin{figure}[h]
    \includegraphics[width=0.96\linewidth]{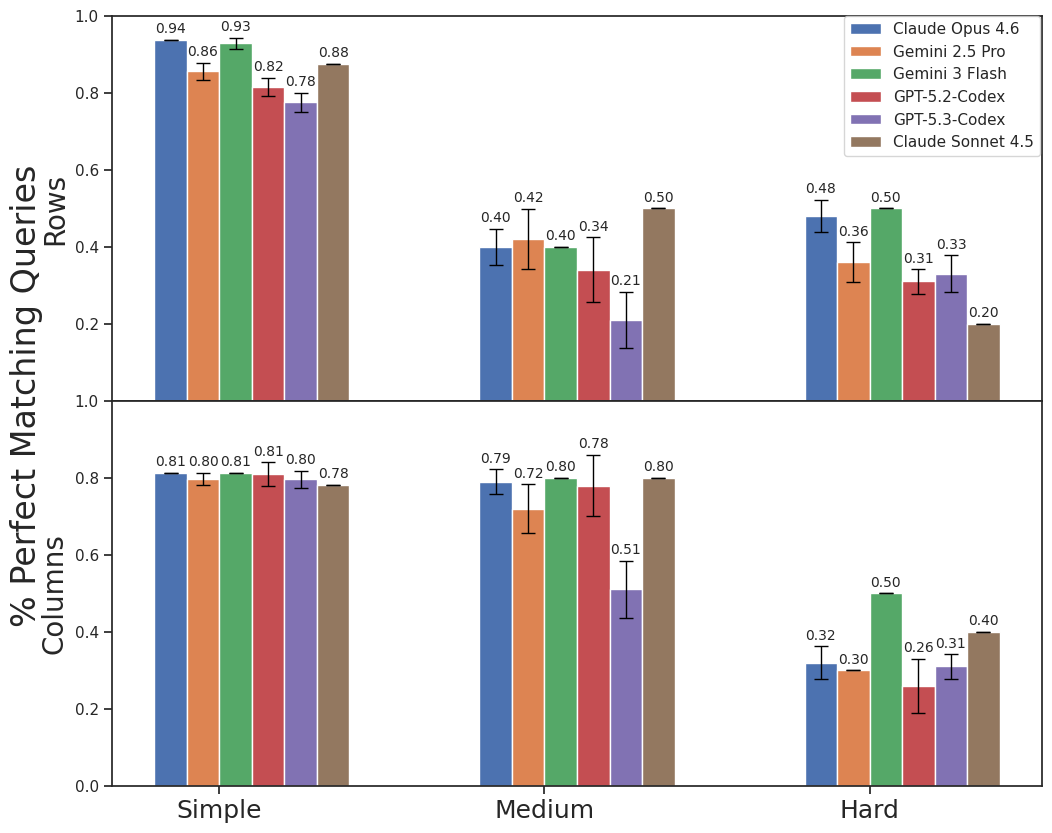}%
    \caption{Performance comparison of the top six LLMs when using the direct method, as a function of the level of query difficulty.}%
    \label{fig:llm_effect_direct}%
\end{figure}

\begin{figure}[h]
   \includegraphics[width=0.96\linewidth]{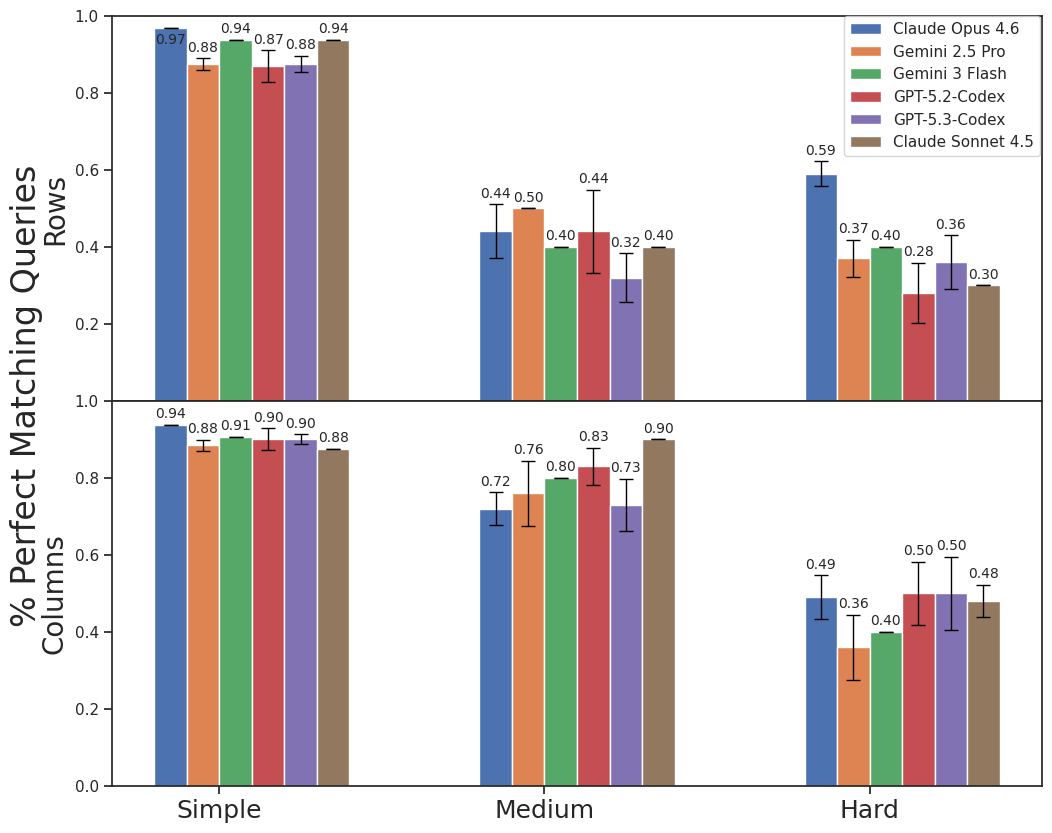}%
   \caption{Performance comparison of the top six LLMs when using the step-by-step method, as a function of the level of query difficulty.}%
   \label{fig:llm_effect}%
\end{figure}

Figure ~\ref{fig:llm_effect_direct} shows the performance of the top six LLMs using the direct method, in terms of the percentage of PM queries for row and column identifiers, as a function of the degree of difficulty of the query. Likewise, Fig. \ref{fig:llm_effect} shows the performance of the top six LLMs using the step-by-step method.
Figs. \ref{fig:claude_llm_comparison}, \ref{fig:gemini_llm_comparison}, and \ref{fig:gpt_llm_comparison} in Appendix \ref{additional results} show the progression over time of the LLM performance according to each provider for the models evaluated (Claude, Gemini and GPT) for our T2S task.
The results of using GPT-5.2-Codex are shown in Fig. \ref{fig:nosc_vs_sc_gpt52codex}.

\subsection{Analysis of errors} \label{errors}

Three main kinds of execution errors can be distinguished. A timeout error occurs when a query exceeds ALeRCE's 2-minute execution limit, e.g., when the model omits a specific condition needed to limit the volume of the data requested. Alternatively, the model may misinterpret the request, producing an incorrect condition, or place a condition in the wrong scope (e.g., inside a subquery when it should be applied at the outer level). The second category includes undefined objects, tables, relations, columns, and so on~--e.g., names of tables or columns that are not in the database. The third category includes syntax and other errors, such as: datatype mismatch (e.g., UNION types and double precision cannot be matched); cardinality violation (e.g., more than one row returned by a subquery used as an expression); and ambiguous column (e.g., selecting the same column name from two tables without specifying the table reference). 

An error that we found during our system's development was the LLMs' tendency to rename columns. For example, in Table \ref{tab:example_parsing}, the golden column names `dist’ and  `oid\_catalog’ were changed to the predicted columns ‘distance’ and `allwise\_oid\_catalog’, respectively. The first error is semantically and logically correct, but it does not match the real name. We solved this problem by parsing the queries after execution using the Python library ``sqlparse''.

\begin{table}[h]
\caption{Example of parsing columns.}
    \label{tab:example_parsing}
    \begin{tabular}{c l}
        \hline \hline
        \begin{minipage}{1.8cm}
        \textbf{Predicted \ \ \ SQL-Query}
        \end{minipage}
                & \begin{minipage}{6.2cm}
        \begin{lstlisting}[language=SQL]
 SELECT 
    object.oid AS ztf_oid, 
    allwise.oid_catalog AS allwise_oid_catalog, 
    xmatch.dist AS distance ... 
        \end{lstlisting} 
        \end{minipage}
        \\ \\
        \hline
        \begin{minipage}{1.8cm}
        \textbf{Extracted aliased columns}
        \end{minipage}%
                & \begin{minipage}{6.2cm}
                \ \\ \ \\ ‘ztf\_oid’, ‘allwise\_oid\_catalog’, ‘distance’, ... \\
        \end{minipage}%
        \\
        \hline
        \begin{minipage}{1.8cm}
        \textbf{Extracted real columns}
        \end{minipage}
        & \begin{minipage}{6.2cm}
\ \\ \ \\ ‘oid’, ‘oid\_catalog’, ‘dist’, ... \\ \ \\ 
        \end{minipage}
        \\
        \hline
        \begin{minipage}{1.8cm}
        \textbf{Gold columns}
        \end{minipage}
        & \begin{minipage}{6.2cm}
\ \\ \ \\ `oid', `oid\_catalog’, ‘dist’, … \\ \ \\
        \end{minipage}
        \\
        \hline
    \end{tabular}
    
\end{table}

\begin{figure}[t]
    \centering
    \includegraphics[width=1.0\linewidth]{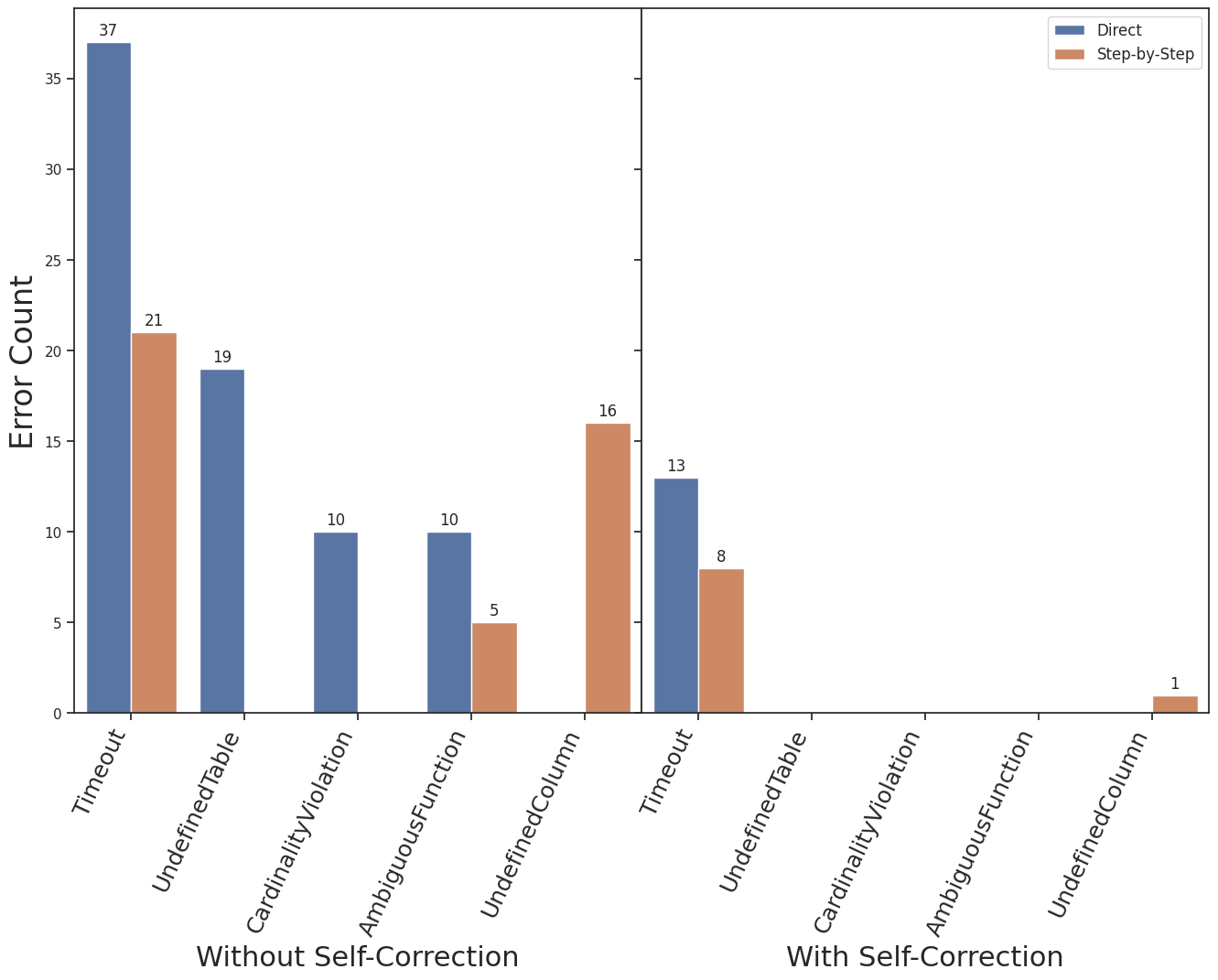}
    \caption{Number and type of execution errors using  Claude Opus 4.6, a) without self-correction and b) with self-correction, for the direct and step-by-step query generation methods.}
    \label{fig:error_type_dir_vs_sbs_claude46opus}
\end{figure}

Figure~\ref{fig:error_type_dir_vs_sbs_claude46opus} compares the execution errors obtained w/ and w/o self-correction when using Claude Opus 4.6. The vast majority of errors are due to timeout. Without self-correction, there are errors associated with undefined columns and tables, cardinality violation, and invalid column references. In Figure~\ref{fig:error_type_dir_vs_sbs_claude46opus} all non-timeout errors are corrected by self-correction, except one undefined column.

\begin{figure*}[t]
    \centering
    \includegraphics[width=1.0\linewidth]{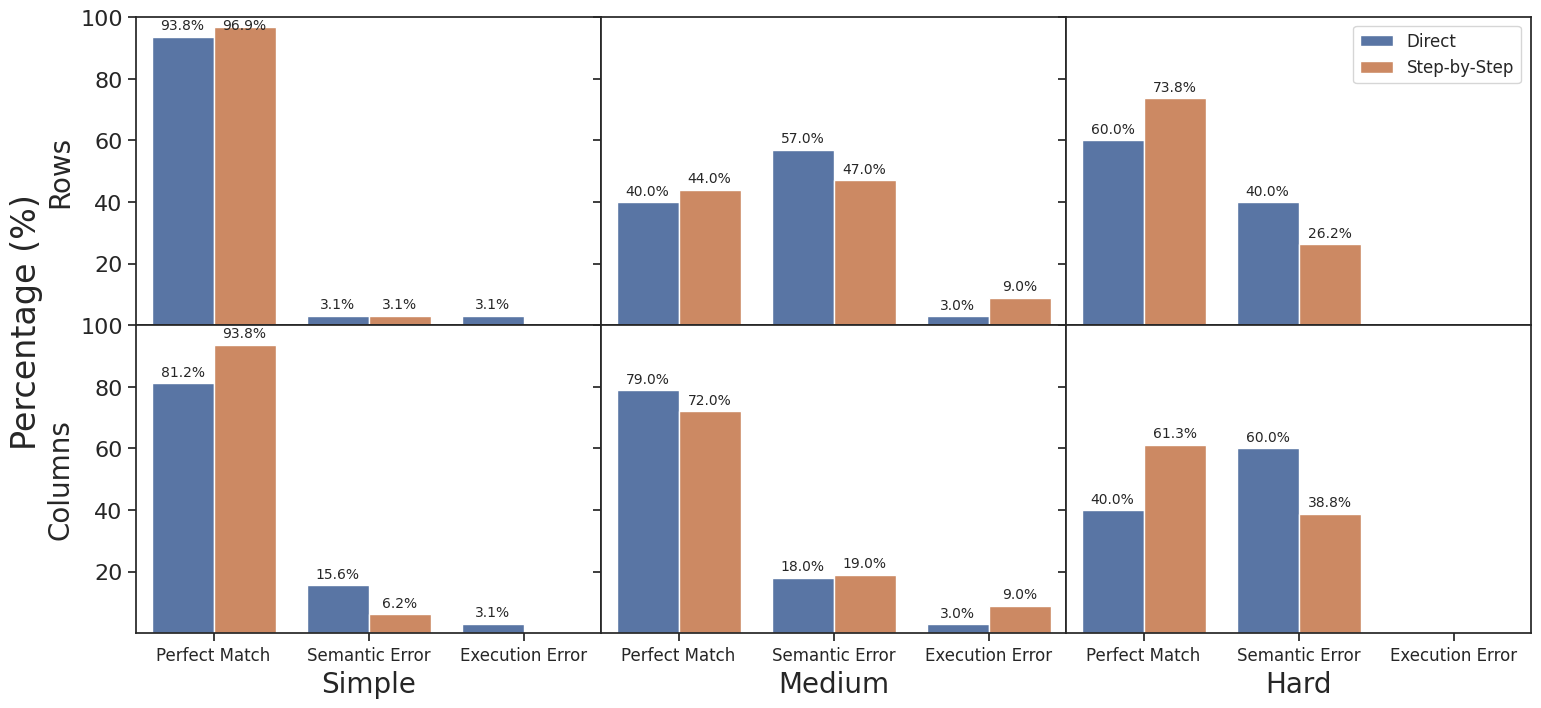}
    \caption{Percentage of perfect queries, semantic errors, and execution errors for the direct and step-by-step query generation methods using Claude Opus 4.6.}
    \label{fig:percentage_errors_dir_vs_sbs_claude46opus}
\end{figure*}

In addition to execution errors, there are semantic errors: the query is executable but does not, or does not always, produce the intended results. Semantic errors encompass a broad range of issues, among them logical errors (e.g., confusing similarly named columns), language comprehension, ambiguity issues and cases where the model lacks specialized domain knowledge. These kinds of errors are not corrected by self-correction. Fig. \ref{fig:percentage_errors_dir_vs_sbs_claude46opus} shows that when using the step-by-step method with Claude Opus 4.6, semantic errors are generally reduced or kept similar to those of the direct method. It can be observed that for medium queries the percentage of semantic errors is much higher for rows than for columns. This discrepancy appears because column evaluation is much simpler than row evaluation. Row precision and recall require the model to satisfy multiple conditions simultaneously, including correctly identifying join paths, applying filters, and constructing appropriate WHERE clauses. Column selection, on the other hand, can be viewed as a classification or association task; the LLM must identify which attributes are relevant to the query and ensure that the generated SQL executes successfully against the database schema. At the medium level, models can still perform the column association reliably but begin to struggle with the more compositional reasoning that row-level accuracy demands.
The analysis of errors for GPT-5.2-Codex and Gemini 3 Flash are shown in Figs. \ref{fig:percentage_errors_dir_vs_sbs_gpt52codex}, and  \ref{fig:count_errors_dir_vs_sbs_gemini3flash} in Appendix \ref{additional results}.

\subsection{Few-shot prompting}

A few-shot prompt for text-to-SQL is a prompting strategy where an LLM is given a small number of example NL/SQL pairs before asking it to generate SQL for a new question. We tested Claude Opus 4.6 and Gemini 2.5 Pro with few-shot prompting. There are three main strategies for example selection in the literature: random, it randomly samples k examples from the available candidates; question similarity selection \citep[QTS;][]{liu2022makes}, it chooses k examples with the most similar questions; masked question similarity selection \citep[MQS;][]{guo2023case}, it replaces table names, column names, and values in all questions with a mask token, and computes the similarities of the embeddings with the k-Nearest Neighbor (kNN) algorithm. We made a few changes in the QTS and MQS strategies, in order to adapt them to our dataset. For both strategies we use a similarity search through embeddings, cosine similarity and top-k selection. For MQS we mask the following elements: ZTF oids, numerical values, month names, and ALeRCE classes.
Table
\ref{tab:few_shot_comparison_by_difficulty_direct} shows the results of 1-shot and 3-shot prompting when using Gemini 2.5 Pro (direct). The best results are obtained with the MQS strategy using 3-shot, which achieves a statistically significant increase for all types of queries, as well as row/column ids, with respect to the 0-shot. Notably the row (column) id performance increased 18 (14) points. QTS 3-shot also achieves statistically significant differences with 0-shot, except for a draw in hard rows. Tables \ref{tab:few_shot_comparison_by_difficulty_sbs_gemini25pro} and \ref{tab:few_shot_comparison_by_difficulty_sbs_claudeopus46} in Appendix \ref{appendix:few_shot} show the results of few-shot prompting for the Gemini 2.5 Pro (sbs) and Claude Opus 4.6 (sbs), respectively. Listing \ref{lst:fewshot_example} shows an example of few-shot prompt.

\begin{table*}[htbp]
\centering
\caption{Few-shot PM results by query difficulty when using Gemini 2.5 Pro (direct).}
\label{tab:few_shot_comparison_by_difficulty_direct}
\begin{tabular}{llcccccc}
\toprule
\toprule
Few-shot & Selection & \multicolumn{6}{c}{Gemini 2.5 Pro} \\
\cmidrule(lr){3-8}
 &  & \multicolumn{2}{c}{Simple} & \multicolumn{2}{c}{Medium} & \multicolumn{2}{c}{Hard} \\
\cmidrule(lr){3-4} \cmidrule(lr){5-6} \cmidrule(lr){7-8}
 &  & Rows & Columns & Rows & Columns & Rows & Columns \\
\midrule
0-shot & - & 0.86 $\pm$ 0.022 & 0.80 $\pm$ 0.016 & 0.42 $\pm$ 0.079 & 0.72 $\pm$ 0.063 & 0.36 $\pm$ 0.052 & 0.30 $\pm$ 0.000 \\
\midrule
1-shot & Random & 0.86 $\pm$ 0.017 $\sim$ & 0.82 $\pm$ 0.026 $\sim$ & 0.50 $\pm$ 0.000 $\uparrow$ & 0.84 $\pm$ 0.055 $\uparrow$ & 0.34 $\pm$ 0.055 $\sim$ & 0.34 $\pm$ 0.114 $\sim$ \\
 & QTS & 0.90 $\pm$ 0.015 $\uparrow$ & 0.85 $\pm$ 0.023 $\uparrow$ & 0.42 $\pm$ 0.042 $\sim$ & 0.79 $\pm$ 0.057 $\uparrow$ & 0.42 $\pm$ 0.063 $\sim$ & 0.27 $\pm$ 0.048 $\sim$ \\
 & MQS & 0.94 $\pm$ 0.022 $\uparrow$ & 0.89 $\pm$ 0.028 $\uparrow$ & 0.38 $\pm$ 0.045 $\sim$ & 0.76 $\pm$ 0.055 $\sim$ & \textbf{0.58 $\pm$ 0.045 $\uparrow$} & 0.36 $\pm$ 0.055 $\sim$ \\
\midrule
3-shot & Random & 0.92 $\pm$ 0.017 $\uparrow$ & 0.87 $\pm$ 0.014 $\uparrow$ & 0.44 $\pm$ 0.055 $\sim$ & 0.80 $\pm$ 0.071 $\uparrow$ & 0.34 $\pm$ 0.055 $\sim$ & 0.38 $\pm$ 0.045 $\uparrow$ \\
 & QTS & \textbf{0.97 $\pm$ 0.000 $\uparrow$} & \textbf{0.93 $\pm$ 0.017 $\uparrow$} & 0.58 $\pm$ 0.045 $\uparrow$ & 0.80 $\pm$ 0.000 $\uparrow$ & 0.40 $\pm$ 0.100 $\sim$ & \textbf{0.40 $\pm$ 0.000 $\uparrow$} \\
 & MQS & 0.94 $\pm$ 0.000 $\uparrow$ & 0.86 $\pm$ 0.017 $\uparrow$ & \textbf{0.60 $\pm$ 0.000 $\uparrow$} & \textbf{0.86 $\pm$ 0.055 $\uparrow$} & 0.48 $\pm$ 0.045 $\uparrow$ & \textbf{0.40 $\pm$ 0.000 $\uparrow$} \\
\bottomrule
\end{tabular}
\tablefoot{The best value per table column is bolded. Arrows compare the few-shot setting against the 0-shot baseline for each query difficulty using two-sided permutation tests on perfect-match scores grouped by run; $\uparrow$/$\downarrow$ indicate significant higher/lower results at $\alpha=0.05$ with no multiple-testing correction, and $\sim$ indicates not significant.}
\end{table*}

\section{Discussion}
\label{sec:discussion}

In this section we discuss two possible ways to improve our proposed framework, especially from the viewpoint of putting in production our T2S system in the near future: structured outputs and function/tool calling.
Regarding specialized tools, function calling can considerably expand the model’s capabilities. In our NL/SQL dataset, the external knowledge associated with some queries, which is mainly composed of MJD dates and celestial coordinates (e.g., RA/Dec), was added by hand. However, this kind of information can be obtained automatically through function/tool calling. These tools leverage the model’s ability to interpret the user’s request and format the appropriate input for each function. Listing \ref{lst:specialized_tool_example} shows an example of a tool that converts dates found in the user request to MJD format. Table \ref{tab:combined_vertical_results_by_function_calling} (see Appendix \ref{appendix:func_call} )  shows the PM results when using Claude Opus 4.6, Gemini 2.5 Pro, and Gemini 3 Flash w/ and w/o the function calling for MJD. It can be observed that the performance increases or remains the same for the sbs query generation method, except for hard rows with Claude Opus 4.6. The main point here is that the function/tool calling simplifies the operation from the user viewpoint, and therefore it is highly recommended to include it in production.

Using SQL as a native structured output means the LLM generates executable database queries in a formally constrained syntax. This is structured output because SQL grammar is formal, tables/columns have schema constraints, and syntax validity is useful. For text-to-SQL tasks, Claude, Gemini, and GPT models all support structured output, but they implement it differently at the API and decoding levels. To obtain structured outputs, we used the integrated structured outputs feature of the API providers. OpenAI\footnote{https://developers.openai.com/api/docs/guides/structured-outputs}, Anthropic\footnote{https://platform.claude.com/docs/en/build-with-claude/structured-outputs}, and Google\footnote{https://ai.google.dev/gemini-api/docs/structured-output} have a specific variable named ``text\_format'', ``output\_format'', and ``response\_format'', respectively, where a JSON schema structure can be passed. Another option is to define a Pydantic BaseModel structure, as shown in all the model guide pages, and selecting the `` ${\rm extra}=`forbid'$ '' configuration to enable strict mode to comply with strict JSON schema requirements and matching. Some benefits of structured outputs include reliable type-safety, seamless system integration, reduced hallucinations, and simpler prompting to obtain consistent formatting.
We tested a type of structured output for the decomposition stage of the step-by-step method for medium/hard queries, defined in Pydantic as shown in Listing \ref{lst:sbs_plan_pydantic} (see Appendix \ref{appendix:func_call}) and that can be transformed to a JSON schema. Listing \ref{lst:sbs_plan_json} shows the Pydantic structure transformed into a JSON dictionary schema that is the format given to the model, forcing the latter to return a list of steps to build the required SQL, and then joining all these steps into one large query in the SQL generation stage.
We tested this format with the medium and hard queries. Table \ref{tab:structured_plan_comparison_by_difficulty} shows the performance with the normal output and the structured step-by-step output for Gemini 3 Flash, Gemini 2.5 Pro, and Claude Opus 4.6. The models did not show a significant performance improvement. Gemini 3 Flash obtained better results for columns for medium queries, while for rows in medium queries, a degraded performance is obtained for Gemini 3 Flash and Claude Opus 4.6. Nevertheless, some cases showed a more consistent output with more successful runs, reducing execution errors. This approach is worth to be explored in more detail in future work.

A natural extension of the current self-correction module would be to constrain its output via a typed schema (e.g., a Pydantic-validated correction object carrying identifier-replacement records restricted to the database vocabulary, applied deterministically on the SQL Abstract Syntax Tree (AST)), with a free-form fallback for syntax, type, and structural errors. This would prevent hallucinated table and column names in the most common error class while preserving the flexibility required for user-requested aliases and complex query structures. This is a valuable direction for future work and we plan to explore it, particularly the post-generation validation approach, which is the most compatible with our current API-based pipeline.

\section{Conclusions}
\label{sec:conclusions}

In this work, we proposed a framework for performing the text-to-SQL parsing task in astronomy using in-context learning with LLMs. With this aim, we built a dataset of NL/SQL queries for the ALeRCE database. We show that the proposed framework outperforms the direct inference method and that self-correction, in general, improves results. In the performance comparison of thirteen LLMs, we found that Claude Opus 4.6, Gemini 2.5 Pro, Gemini 3 Flash and GPT 5.2 Codex are the best LLMs for the task at hand. In future work, we plan to explore technologies beyond classic prompt engineering, such as Retrieval-Augmented Generation (RAG) and fine-tuning LLMs on generating SQL outputs. Future work will evaluate the proposed framework against both specialized text-to-SQL systems, e.g., SQLCoder\footnote{https://github.com/defog-ai/sqlcoder}, and modern open-weight reasoning models such as DeepSeek, Qwen, and LLaMA variants. In addition, developing a more comprehensive self-review process that evaluates semantic correctness, not only execution errors, is a promising direction for future work.  Our ultimate goal is to implement an online virtual assistant for ALeRCE users in the Rubin era.

\begin{acknowledgements}
This work gratefully acknowledges funding from ANID: Millennium Science Initiative, AIM23-0001 (P.A.E, J.E-M, F.F., G.C.-V. A.M.M.A, G.P., F.E.B., M.C., R.D.); ANID/Basal (CATA) grant FB210003 (F.F., M.C., F.E.B.); and FONDECYT Regular grants 1220829 (P.A.E.),  1231877 (G.C.V), 1241005 (F.E.B.), and 1231637 (M.C.). This work was supported by Centro de Modelamiento Matemático (CMM) BASAL fund FB210005 from ANID-Chile. A.B acknowledges support from the Deutsche Forschungsgemeinschaft (DFG, German Research Foundation) under Germany's Excellence Strategy – EXC 2094 – 390783311. We acknowledge Zhen Guo and Arti Joshi for their contribution in providing samples for the dataset used in this paper.
\end{acknowledgements}

\bibliographystyle{aa}
\bibliography{library}{}

\onecolumn

\begin{appendix}

\clearpage

\section{\\ ALeRCE database}
\label{appendix:alerce_db}

\begin{figure}[h]
\centering
    \includegraphics[width=0.8\linewidth]{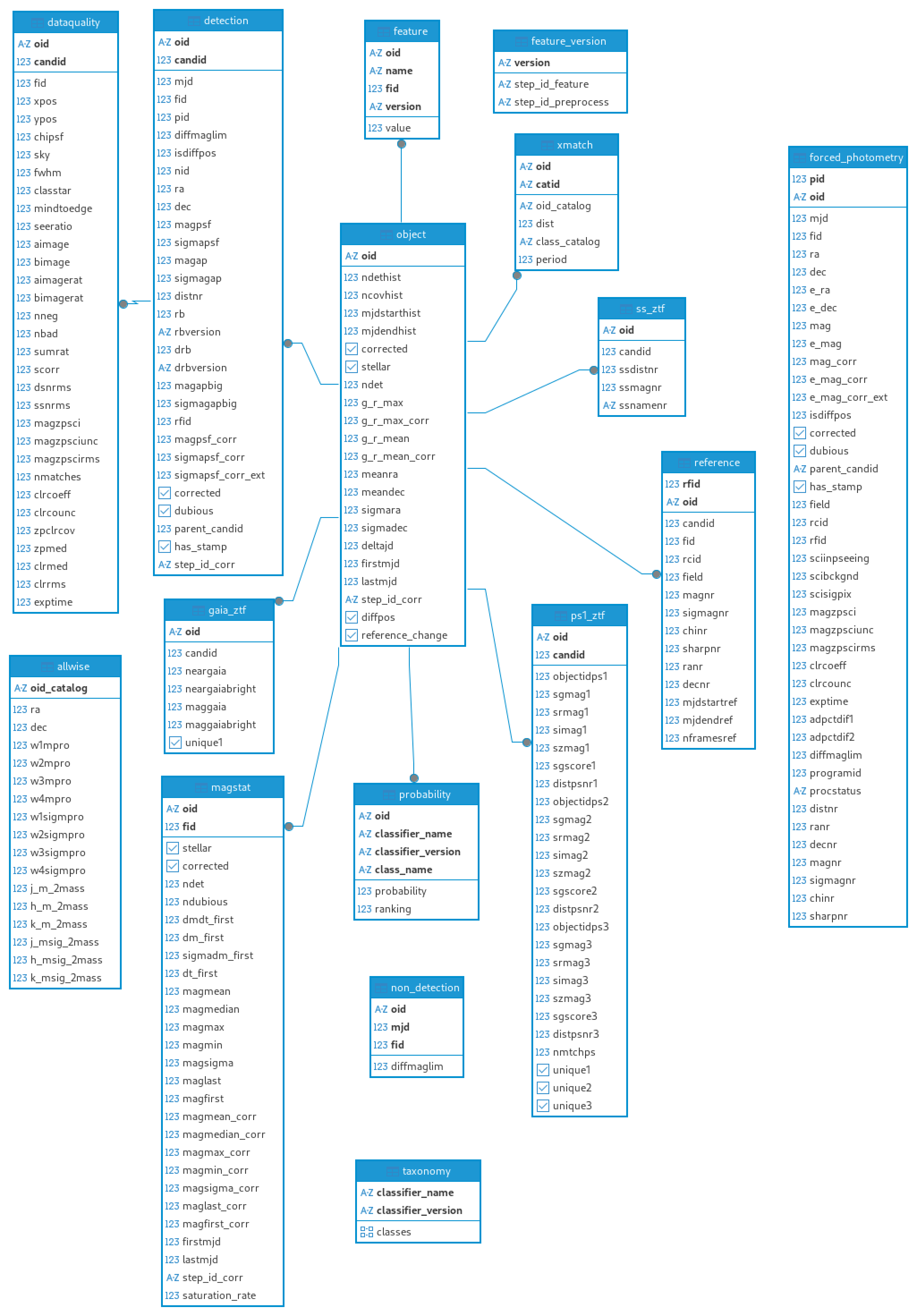}
    \caption{Entity-relationship diagram of the ALeRCE database.}
    \label{fig:alerce_ER_diagram}
\end{figure}

\clearpage
\newpage

\section{Examples of queries}
\label{appendix:examples_diff}
\renewcommand{\thelstlisting}{B.\arabic{lstlisting}}

\begin{table}[h]
\caption{Example of a simple query.}
\label{tab:example_simple_query}
\centering
\small
\begin{tabular}{l p{15cm}l}
\hline \hline
Request &
  For objects with ZTF identifiers 'ZTF21aaobkmg', 'ZTF21aaomuka', find all rows in the 'probability' table and light curve classifier that have ranking 1 or 2. Return all columns from such table, sort by ranking \\ \hline
\begin{minipage}{2cm}
Gold \\ SQL Query 
\end{minipage}
&
\begin{minipage}{15cm}
 \begin{lstlisting}[language=SQL]
 SELECT 
    *
 FROM 
    probability 
 WHERE 
    oid IN ('ZTF21aaobkmg','ZTF21aaomuka') AND 
    classifier_name = 'lc_classifier' AND 
    ranking <= 2 
 ORDER BY ranking
\end{lstlisting} 
\end{minipage}
 \\ \hline

\end{tabular}
\end{table}

\begin{table}[h]
\caption{Example of a medium query.}
\label{tab:example_medium_query}
\centering
\small
\begin{tabular}{l p{15cm}l}
\hline \hline
Request &
  For Solar System identifiers '2003FP134' and '2009UK56', get all detections for all ZTF objects that lie within 2 arcsec from any of them. Return the following columns, sort by MPC name and detection date: all columns from the 'ss\_ztf' table; and detection date, filter identifier, isdiffpos flag, RA Dec coordinates, and difference magnitude (and its uncertainty) \\ \hline
  \begin{minipage}{2cm}
Gold \\ SQL Query 
\end{minipage} 
& 
\begin{minipage}{15cm}
 \begin{lstlisting}[language=SQL]
SELECT 
    ss_ztf.*, detection.mjd, detection.fid, detection.isdiffpos, detection.ra, detection.dec, detection.magpsf, detection.sigmapsf 
FROM 
    ss_ztf 
INNER JOIN detection ON ss_ztf.oid = detection.oid AND ss_ztf.candid = detection.candid 
WHERE 
    ssnamenr IN ('2003FP134','2009UK56') AND 
    ssdistnr <2
ORDER BY ssnamenr, mjd
\end{lstlisting} 
\end{minipage}
  \\ \hline
\end{tabular}
\end{table}

\begin{table}[!htbp]
\caption{Example of a hard query.}
\label{tab:example_adv_query}
\centering
\small
\begin{tabular}{l p{15cm}l}
\hline \hline
Request &
  Find all detections, non-detections and forced photometry points for the ZTF object 'ZTF24aamtvxb'. Return all epochs in the same output table, including the following columns: ZTF identifier, epoch date, filter identifier, isdiffpos flag, detection difference magnitude and its uncertainty, 5-sigma magnitude limit, forced difference magnitude and its uncertainty, and a column named 'table' (that contains the name of the table of origin for each epoch) \\ \hline
Gold SQL Query &
\begin{minipage}{15.1cm}
 \begin{lstlisting}[language=SQL]
 SELECT
    oid, mjd, fid, isdiffpos, magpsf, sigmapsf, NULL as diffmaglim, CAST(NULL AS bigint) as mag, CAST(NULL AS bigint) as e_mag, 'detection' as table
FROM 
    detection
WHERE 
    oid = 'ZTF24aamtvxb' UNION ALL SELECT oid, mjd, fid, NULL as isdiffpos, NULL as magpsf, NULL as sigmapsf, diffmaglim, CAST(NULL AS bigint) as mag, CAST(NULL AS bigint) as e_mag, 'non_detection' as table 
FROM
    non_detection WHERE oid = 'ZTF24aamtvxb' 
UNION ALL 
SELECT 
    oid, mjd, fid, isdiffpos, NULL as magpsf, NULL as sigmapsf, NULL as diffmaglim, mag, e_mag, 'forced_photometry' as table 
FROM 
    forced_photometry 
WHERE 
    oid = 'ZTF24aamtvxb'
 \end{lstlisting} 
\end{minipage}
 \\ \hline
\end{tabular}
\end{table}

\section{Prompt formats} \label{appendix:self_corr_prompts}
\renewcommand{\thelstlisting}{C.\arabic{lstlisting}}

\begin{lstlisting}[language={python}, caption={Schema linking prompt. }, label={lst:schema_linking_prompt}]
# Given the user request, select the tables needed to generate a SQL query
# The Database has the following tables:

# TABLE "object": contains basic filter and time-aggregated statistics such as location, number of observations, and the times of first and last detection.
...

# Give ONLY the TABLES that are needed to generate the SQL query.
# Give the answer in the following format: ['table1', 'table2', 'table3', ...]. For example, if the TABLES needed for the user request are TABLE object and TABLE taxonomy, then you should type: ['object', 'taxonomy']
# Just give the tables and ignore any other task given in the request given as "request".
# Remember to use the exact name of the TABLES, as they are written in the DATABASE SCHEMA. Do NOT create table names.
# If you think that no table mentioned above is needed, then type: ""
# Request: 
[{User Request}]
\end{lstlisting}

\begin{lstlisting}[language={python}, caption={Difficulty classification prompt. }, label={lst:diff_Class_prompt}]
# For the given request, classify it by difficulty as "simple", "medium", or "hard" based on the next description.

# Simple label description
# Medium label description
# Hard label description

# Database schema. It is not necessary to use all the tables in the schema, but you can use them if you need them.
[{Table_Schema}]
[{Final instructions}]
# Request: 
[{User Request}]
\end{lstlisting}

\begin{lstlisting}[language={python}, caption={Example of timeout self-correction prompt.}, label={lst:timeout_selfcorr_prompt}]
[{self_correction_task}]
# Correct a SQL query given the next user request:
[{user_request}]
# These are the table schemas. Assume that only the following tables are required for the query:
[{table_schemas}]
# The following query is not working due to a timeout error, correct the query using the correct database schema or nested queries to optimize.
# SQL Query
[{sql_pred}]
# Error returned when executing the query in the ALeRCE database
[{sql_error}]

# Follow these guidelines to correct the query:
# - Check if the SQL code includes the necessary conditions to optimize the query, and if the query is using the correct database schema or nested queries to optimize.
#     - It is possible that the query is too complex and that nested queries are necessary to optimize it.
#     - If there is a JOIN or a sub-query between some table and probability, check if the condition 'ranking=1' is set in the probability table, unless the request said otherwise.
# - If there are conditions involving dates or times, check if the dates are not too far away, or are in a reasonable range.
# - If the probability table is used, always use the next conditions, unless the user explicitly specifies different probability conditions.
# - Ensure there are at least 3 conditions on the probability table, because if not, the query is too general. Add more conditions if necessary.

# Check the query and correct it by modifying the SQL code where the error is found.
# Add COMMENTS IN PostgreSQL format so that the user can understand.
# Answer ONLY with the SQL query
# SQL:
\end{lstlisting}

\section{Perfect match rates by LLM} \label{appendix:pm_rates}

\begin{table*}[h!]
\footnotesize
\centering
\setlength{\tabcolsep}{4pt}
\caption{Perfect match rates by LLM, query generation method, query difficulty, and w/ or w/o self-correction.}
\begin{tabular}{lllllll}
\toprule
\toprule
 & \multicolumn{2}{c}{Simple} & \multicolumn{2}{c}{Medium} & \multicolumn{2}{c}{Hard} \\
\cmidrule(lr){2-3} \cmidrule(lr){4-5} \cmidrule(lr){6-7}
Model & w/o Self-Corr & w/ Self-Corr & w/o Self-Corr & w/ Self-Corr & w/o Self-Corr & w/ Self-Corr \\
\midrule
\multicolumn{7}{l}{\textbf{ID Match (Rows) -- Direct}} \\
Claude 3.7 & 0.844 $\pm$ 0.0 & 0.866 $\pm$ 0.021 & 0.160 $\pm$ 0.084 & 0.160 $\pm$ 0.084 & 0.280 $\pm$ 0.079 & 0.380 $\pm$ 0.079 \\
Claude Opus 4.6 & 0.844 $\pm$ 0.0 & \textbf{0.938 $\pm$ 0.0} & 0.250 $\pm$ 0.053 & 0.400 $\pm$ 0.047 & \textbf{0.400 $\pm$ 0.0} & \textbf{0.480 $\pm$ 0.042} \\
Claude Sonnet 4.5 & 0.844 $\pm$ 0.0 & 0.875 $\pm$ 0.0 & \textbf{0.400 $\pm$ 0.0} & \textbf{0.500 $\pm$ 0.0} & 0.200 $\pm$ 0.0 & 0.200 $\pm$ 0.0 \\
Gemini 2.5 Flash & 0.775 $\pm$ 0.020 & 0.819 $\pm$ 0.025 & \textbf{0.410 $\pm$ 0.057} & 0.410 $\pm$ 0.057 & 0.240 $\pm$ 0.052 & 0.300 $\pm$ 5.9e-17 \\
Gemini 2.5 Pro & 0.809 $\pm$ 0.023 & 0.856 $\pm$ 0.022 & \textbf{0.360 $\pm$ 0.117} & 0.420 $\pm$ 0.079 & 0.160 $\pm$ 0.070 & 0.360 $\pm$ 0.052 \\
Gemini 3 Flash & \textbf{0.866 $\pm$ 0.015} & \textbf{0.928 $\pm$ 0.015} & 0.300 $\pm$ 5.9e-17 & 0.400 $\pm$ 0.0 & \textbf{0.400 $\pm$ 0.0} & \textbf{0.500 $\pm$ 0.0} \\
Gemini 3.1 Flash & 0.844 $\pm$ 0.0 & 0.906 $\pm$ 0.0 & \textbf{0.400 $\pm$ 0.0} & 0.400 $\pm$ 0.0 & 0.200 $\pm$ 0.0 & 0.300 $\pm$ 5.9e-17 \\
GPT-4.1 & 0.753 $\pm$ 0.023 & 0.806 $\pm$ 0.029 & 0.210 $\pm$ 0.088 & 0.260 $\pm$ 0.070 & 0.270 $\pm$ 0.048 & 0.340 $\pm$ 0.052 \\
GPT-4o & \textbf{0.856 $\pm$ 0.016} & 0.891 $\pm$ 0.022 & 0.130 $\pm$ 0.048 & 0.260 $\pm$ 0.084 & 0.0 $\pm$ 0.0 & 0.110 $\pm$ 0.032 \\
GPT-5 & 0.787 $\pm$ 0.032 & 0.822 $\pm$ 0.033 & \textbf{0.350 $\pm$ 0.085} & 0.380 $\pm$ 0.063 & 0.230 $\pm$ 0.048 & 0.290 $\pm$ 0.032 \\
GPT-5.2 & 0.784 $\pm$ 0.023 & 0.800 $\pm$ 0.026 & 0.280 $\pm$ 0.063 & 0.300 $\pm$ 0.067 & \textbf{0.370 $\pm$ 0.116} & 0.380 $\pm$ 0.132 \\
GPT-5.2-Codex & 0.731 $\pm$ 0.030 & 0.816 $\pm$ 0.023 & \textbf{0.320 $\pm$ 0.092} & 0.340 $\pm$ 0.084 & 0.280 $\pm$ 0.063 & 0.310 $\pm$ 0.032 \\
GPT-5.3-Codex & 0.753 $\pm$ 0.018 & 0.775 $\pm$ 0.025 & 0.200 $\pm$ 0.067 & 0.210 $\pm$ 0.074 & 0.190 $\pm$ 0.032 & 0.330 $\pm$ 0.048 \\
\cmidrule(lr){1-7}
\multicolumn{7}{l}{\textbf{ID Match (Rows) -- Step-by-Step}} \\
Claude 3.7 & 0.794 $\pm$ 0.016 & 0.828 $\pm$ 0.030 & 0.230 $\pm$ 0.177 & 0.250 $\pm$ 0.165 & 0.150 $\pm$ 0.085 & 0.250 $\pm$ 0.085 \\
Claude Opus 4.6 & \textbf{0.938 $\pm$ 0.0} & \textbf{0.969 $\pm$ 0.0} & 0.380 $\pm$ 0.042 & \textbf{0.440 $\pm$ 0.070} & \textbf{0.530 $\pm$ 0.048} & \textbf{0.590 $\pm$ 0.032} \\
Claude Sonnet 4.5 & 0.906 $\pm$ 0.0 & 0.938 $\pm$ 0.0 & 0.400 $\pm$ 0.0 & 0.400 $\pm$ 0.0 & 0.200 $\pm$ 0.0 & 0.300 $\pm$ 5.9e-17 \\
Gemini 2.5 Flash & 0.853 $\pm$ 0.015 & 0.853 $\pm$ 0.015 & 0.200 $\pm$ 0.0 & 0.250 $\pm$ 0.053 & 0.190 $\pm$ 0.074 & 0.230 $\pm$ 0.048 \\
Gemini 2.5 Pro & 0.847 $\pm$ 0.023 & 0.875 $\pm$ 0.015 & \textbf{0.490 $\pm$ 0.032} & \textbf{0.500 $\pm$ 0.0} & 0.250 $\pm$ 0.053 & 0.370 $\pm$ 0.048 \\
Gemini 3 Flash & \textbf{0.938 $\pm$ 0.0} & 0.938 $\pm$ 0.0 & 0.390 $\pm$ 0.032 & 0.400 $\pm$ 0.0 & 0.300 $\pm$ 5.9e-17 & 0.400 $\pm$ 0.0 \\
Gemini 3.1 Flash & 0.812 $\pm$ 0.0 & 0.844 $\pm$ 0.0 & 0.300 $\pm$ 5.9e-17 & 0.300 $\pm$ 5.9e-17 & 0.100 $\pm$ 0.0 & 0.200 $\pm$ 0.0 \\
GPT-4.1 & 0.797 $\pm$ 0.022 & 0.809 $\pm$ 0.034 & 0.270 $\pm$ 0.082 & 0.300 $\pm$ 0.115 & 0.210 $\pm$ 0.074 & 0.280 $\pm$ 0.079 \\
GPT-4o & 0.875 $\pm$ 0.033 & 0.900 $\pm$ 0.029 & 0.180 $\pm$ 0.042 & 0.270 $\pm$ 0.048 & 0.130 $\pm$ 0.067 & 0.220 $\pm$ 0.063 \\
GPT-5 & 0.856 $\pm$ 0.030 & 0.884 $\pm$ 0.039 & 0.370 $\pm$ 0.048 & 0.390 $\pm$ 0.032 & 0.190 $\pm$ 0.032 & 0.260 $\pm$ 0.070 \\
GPT-5.2 & 0.806 $\pm$ 0.029 & 0.809 $\pm$ 0.031 & 0.280 $\pm$ 0.042 & 0.290 $\pm$ 0.032 & 0.290 $\pm$ 0.074 & 0.290 $\pm$ 0.074 \\
GPT-5.2-Codex & 0.847 $\pm$ 0.045 & 0.869 $\pm$ 0.041 & \textbf{0.420 $\pm$ 0.092} & \textbf{0.440 $\pm$ 0.107} & 0.250 $\pm$ 0.053 & 0.280 $\pm$ 0.079 \\
GPT-5.3-Codex & 0.838 $\pm$ 0.013 & 0.875 $\pm$ 0.021 & 0.310 $\pm$ 0.074 & 0.320 $\pm$ 0.063 & 0.350 $\pm$ 0.053 & 0.360 $\pm$ 0.070 \\
\midrule
\multicolumn{7}{l}{\textbf{Column Match -- Direct}} \\
Claude 3.7 & 0.738 $\pm$ 0.030 & 0.762 $\pm$ 0.034 & 0.680 $\pm$ 0.042 & 0.790 $\pm$ 0.057 & 0.310 $\pm$ 0.088 & 0.280 $\pm$ 0.092 \\
Claude Opus 4.6 & 0.797 $\pm$ 0.016 & \textbf{0.812 $\pm$ 0.0} & 0.640 $\pm$ 0.052 & 0.790 $\pm$ 0.032 & 0.220 $\pm$ 0.042 & 0.220 $\pm$ 0.042 \\
Claude Sonnet 4.5 & 0.781 $\pm$ 0.0 & 0.781 $\pm$ 0.0 & 0.500 $\pm$ 0.0 & 0.700 $\pm$ 0.0 & 0.300 $\pm$ 5.9e-17 & 0.300 $\pm$ 5.9e-17 \\
Gemini 2.5 Flash & 0.775 $\pm$ 0.025 & \textbf{0.800 $\pm$ 0.022} & 0.760 $\pm$ 0.052 & 0.820 $\pm$ 0.079 & 0.240 $\pm$ 0.052 & 0.350 $\pm$ 0.071 \\
Gemini 2.5 Pro & \textbf{0.828 $\pm$ 0.016} & \textbf{0.828 $\pm$ 0.016} & 0.760 $\pm$ 0.097 & 0.820 $\pm$ 0.063 & 0.200 $\pm$ 0.067 & 0.280 $\pm$ 0.042 \\
Gemini 3 Flash & 0.803 $\pm$ 0.015 & \textbf{0.812 $\pm$ 0.0} & 0.800 $\pm$ 0.0 & 0.800 $\pm$ 0.0 & \textbf{0.400 $\pm$ 0.0} & \textbf{0.500 $\pm$ 0.0} \\
Gemini 3.1 Flash & 0.812 $\pm$ 0.0 & \textbf{0.812 $\pm$ 0.0} & \textbf{0.900 $\pm$ 0.0} & \textbf{0.900 $\pm$ 0.0} & 0.300 $\pm$ 5.9e-17 & 0.300 $\pm$ 5.9e-17 \\
GPT-4.1 & 0.709 $\pm$ 0.015 & 0.741 $\pm$ 0.015 & 0.600 $\pm$ 0.047 & 0.650 $\pm$ 0.053 & 0.230 $\pm$ 0.048 & 0.230 $\pm$ 0.048 \\
GPT-4o & 0.728 $\pm$ 0.026 & 0.738 $\pm$ 0.016 & 0.270 $\pm$ 0.082 & 0.450 $\pm$ 0.085 & 0.100 $\pm$ 0.0 & 0.120 $\pm$ 0.042 \\
GPT-5 & 0.750 $\pm$ 0.026 & 0.775 $\pm$ 0.025 & 0.690 $\pm$ 0.110 & 0.760 $\pm$ 0.107 & 0.250 $\pm$ 0.071 & 0.280 $\pm$ 0.063 \\
GPT-5.2 & 0.762 $\pm$ 0.034 & 0.778 $\pm$ 0.031 & 0.750 $\pm$ 0.071 & 0.770 $\pm$ 0.048 & 0.270 $\pm$ 0.067 & 0.300 $\pm$ 0.067 \\
GPT-5.2-Codex & 0.728 $\pm$ 0.033 & \textbf{0.809 $\pm$ 0.031} & 0.760 $\pm$ 0.084 & 0.790 $\pm$ 0.088 & 0.280 $\pm$ 0.042 & 0.320 $\pm$ 0.042 \\
GPT-5.3-Codex & 0.759 $\pm$ 0.030 & 0.797 $\pm$ 0.022 & 0.560 $\pm$ 0.052 & 0.560 $\pm$ 0.052 & 0.190 $\pm$ 0.057 & 0.330 $\pm$ 0.048 \\
\cmidrule(lr){1-7}
\multicolumn{7}{l}{\textbf{Column Match -- Step-by-Step}} \\
Claude 3.7 & 0.734 $\pm$ 0.034 & 0.778 $\pm$ 0.027 & 0.530 $\pm$ 0.149 & 0.660 $\pm$ 0.143 & \textbf{0.280 $\pm$ 0.103} & 0.260 $\pm$ 0.052 \\
Claude Opus 4.6 & \textbf{0.938 $\pm$ 0.0} & \textbf{0.938 $\pm$ 0.0} & 0.650 $\pm$ 0.053 & 0.720 $\pm$ 0.042 & 0.300 $\pm$ 5.9e-17 & \textbf{0.310 $\pm$ 0.032} \\
Claude Sonnet 4.5 & 0.875 $\pm$ 0.0 & 0.875 $\pm$ 0.0 & \textbf{0.800 $\pm$ 0.0} & \textbf{0.900 $\pm$ 0.0} & 0.250 $\pm$ 0.053 & 0.300 $\pm$ 5.9e-17 \\
Gemini 2.5 Flash & 0.906 $\pm$ 0.0 & 0.906 $\pm$ 0.0 & \textbf{0.850 $\pm$ 0.053} & \textbf{0.880 $\pm$ 0.042} & 0.200 $\pm$ 0.0 & 0.240 $\pm$ 0.052 \\
Gemini 2.5 Pro & 0.906 $\pm$ 0.0 & 0.906 $\pm$ 0.0 & \textbf{0.790 $\pm$ 0.032} & 0.800 $\pm$ 0.047 & 0.240 $\pm$ 0.052 & \textbf{0.310 $\pm$ 0.057} \\
Gemini 3 Flash & 0.906 $\pm$ 0.0 & 0.906 $\pm$ 0.0 & \textbf{0.800 $\pm$ 0.0} & 0.810 $\pm$ 0.032 & 0.300 $\pm$ 5.9e-17 & 0.300 $\pm$ 5.9e-17 \\
Gemini 3.1 Flash & 0.875 $\pm$ 0.0 & 0.906 $\pm$ 0.0 & \textbf{0.800 $\pm$ 0.0} & 0.800 $\pm$ 0.0 & 0.300 $\pm$ 5.9e-17 & 0.300 $\pm$ 5.9e-17 \\
GPT-4.1 & 0.903 $\pm$ 0.010 & 0.903 $\pm$ 0.010 & 0.680 $\pm$ 0.063 & 0.750 $\pm$ 0.053 & 0.250 $\pm$ 0.053 & 0.290 $\pm$ 0.032 \\
GPT-4o & 0.822 $\pm$ 0.021 & 0.822 $\pm$ 0.021 & 0.540 $\pm$ 0.097 & 0.700 $\pm$ 0.094 & \textbf{0.360 $\pm$ 0.070} & \textbf{0.390 $\pm$ 0.074} \\
GPT-5 & 0.856 $\pm$ 0.037 & 0.872 $\pm$ 0.031 & \textbf{0.750 $\pm$ 0.097} & 0.830 $\pm$ 0.067 & 0.050 $\pm$ 0.053 & 0.180 $\pm$ 0.092 \\
GPT-5.2 & 0.863 $\pm$ 0.016 & 0.863 $\pm$ 0.016 & 0.760 $\pm$ 0.052 & 0.770 $\pm$ 0.067 & \textbf{0.390 $\pm$ 0.074} & \textbf{0.390 $\pm$ 0.074} \\
GPT-5.2-Codex & 0.887 $\pm$ 0.037 & 0.900 $\pm$ 0.029 & \textbf{0.810 $\pm$ 0.057} & 0.830 $\pm$ 0.048 & \textbf{0.330 $\pm$ 0.067} & \textbf{0.380 $\pm$ 0.042} \\
GPT-5.3-Codex & 0.875 $\pm$ 0.0 & 0.906 $\pm$ 0.0 & 0.690 $\pm$ 0.057 & 0.730 $\pm$ 0.067 & \textbf{0.370 $\pm$ 0.067} & \textbf{0.380 $\pm$ 0.079} \\
\bottomrule
\end{tabular}
\tablefoot{In each of the four sections of the table, the best mean per column is shown in boldface type, together with every LLM model whose run-level scores are not significantly worse than the best under a two-sided unpaired permutation test (Holm-corrected, $\alpha=0.05$).}
\label{tab:combined_vertical_results_by_exp}
\end{table*}

\newpage
\section{Additional results}
\label{additional results}

\nopagebreak

\begin{table}[h!]
\scriptsize
\centering
\setlength{\tabcolsep}{4pt}
\caption{Total and stagewise maximum per-query token counts by LLM model and query generation method.}
\label{tab:max_tokens_by_stage}
\begin{tabular}{lcccccccccccc}
\toprule
\toprule
Model & \multicolumn{6}{c}{Direct} & \multicolumn{6}{c}{Step-by-Step} \\
\cmidrule(lr){2-7} \cmidrule(lr){8-13}
 & Total & Schema & Difficulty & Plan & SQL & Self-Correction & Total & Schema & Difficulty & Plan & SQL & Self-Correction \\
\midrule
GPT-5.2 & \textbf{42\,671} & 802 & \textbf{0} & \textbf{0} & \textbf{15\,363} & \textbf{28\,877} & 19\,357 & 802 & 4\,328 & 8\,554 & 7\,991 & 5\,303 \\
GPT-5.3-Codex & 39\,907 & 801 & \textbf{0} & \textbf{0} & 12\,261 & 27\,044 & 22\,978 & 801 & 4\,263 & 7\,039 & 6\,574 & 6\,712 \\
GPT-5.2-Codex & 22\,106 & 1\,539 & \textbf{0} & \textbf{0} & 13\,740 & 9\,029 & 35\,236 & 1\,539 & 5\,085 & 12\,228 & 10\,697 & 9\,252 \\
GPT-5 & 26\,655 & 3\,919 & \textbf{0} & \textbf{0} & 13\,312 & 11\,232 & \textbf{58\,071} & 3\,919 & \textbf{6\,609} & \textbf{20\,935} & \textbf{16\,627} & \textbf{17\,157} \\
Claude Opus 4.6 & 8\,193 & 1\,122 & \textbf{0} & \textbf{0} & 4\,716 & 4\,135 & 13\,439 & 1\,122 & 3\,648 & 5\,524 & 4\,415 & 8\,098 \\
GPT-4.1 & 20\,738 & 795 & \textbf{0} & \textbf{0} & 8\,659 & 11\,351 & 20\,279 & 951 & 4\,201 & 7\,557 & 6\,878 & 5\,160 \\
Claude Sonnet 3.7 & 11\,303 & 2\,091 & \textbf{0} & \textbf{0} & 1\,211 & 8\,138 & 8\,243 & 2\,091 & 1\,190 & 1\,162 & 1\,153 & 4\,543 \\
Gemini 2.5 Pro & 20\,741 & \textbf{7\,936} & \textbf{0} & \textbf{0} & 6\,826 & 6\,579 & 34\,344 & \textbf{7\,936} & 4\,606 & 8\,128 & 7\,499 & 6\,795 \\
GPT-4o & 9\,509 & 795 & \textbf{0} & \textbf{0} & 6\,632 & 4\,127 & 18\,677 & 795 & 4\,809 & 7\,622 & 4\,647 & 5\,010 \\
Claude Sonnet 4.5 & 5\,737 & 1\,729 & \textbf{0} & \textbf{0} & 947 & 3\,692 & 12\,701 & 1\,729 & 1\,194 & 2\,088 & 1\,101 & 7\,872 \\
Gemini 3 Flash & 8\,557 & 1\,924 & \textbf{0} & \textbf{0} & 6\,084 & 3\,437 & 18\,789 & 1\,924 & 4\,486 & 7\,458 & 6\,696 & 6\,595 \\
Gemini 3.1 Flash & 8\,206 & 913 & \textbf{0} & \textbf{0} & 7\,067 & 3\,397 & 27\,134 & 913 & 5\,144 & 7\,835 & 7\,982 & 5\,349 \\
\bottomrule
\end{tabular}
\tablefoot{Values represent the highest token count observed across all evaluated queries for each stage. The largest value per column is bolded to indicate the upper bound.}
\end{table}

\begin{figure}[h!]
    \centering
    \begin{subfigure}[b]{0.4\textwidth}
        \centering
        \includegraphics[width=1.0\linewidth]{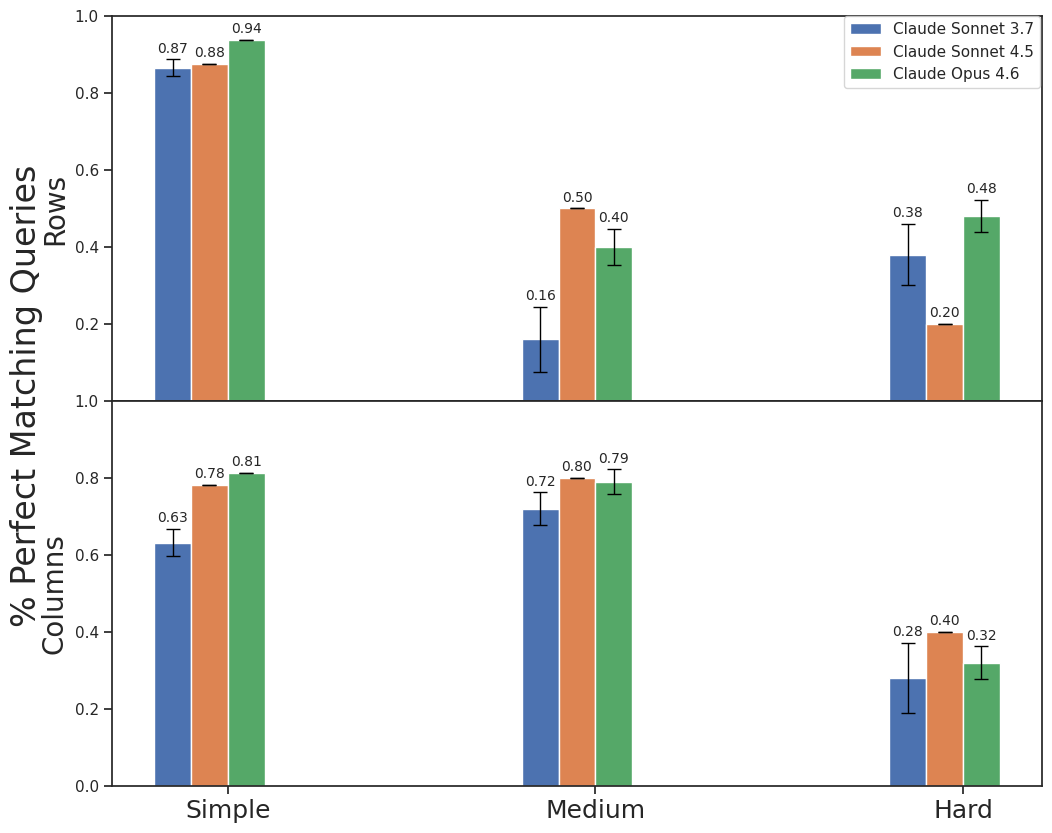}
        \caption{Direct generation method}
    \end{subfigure}%
    \begin{subfigure}[b]{0.4\textwidth}
        \centering
       \includegraphics[width=1.0\linewidth]{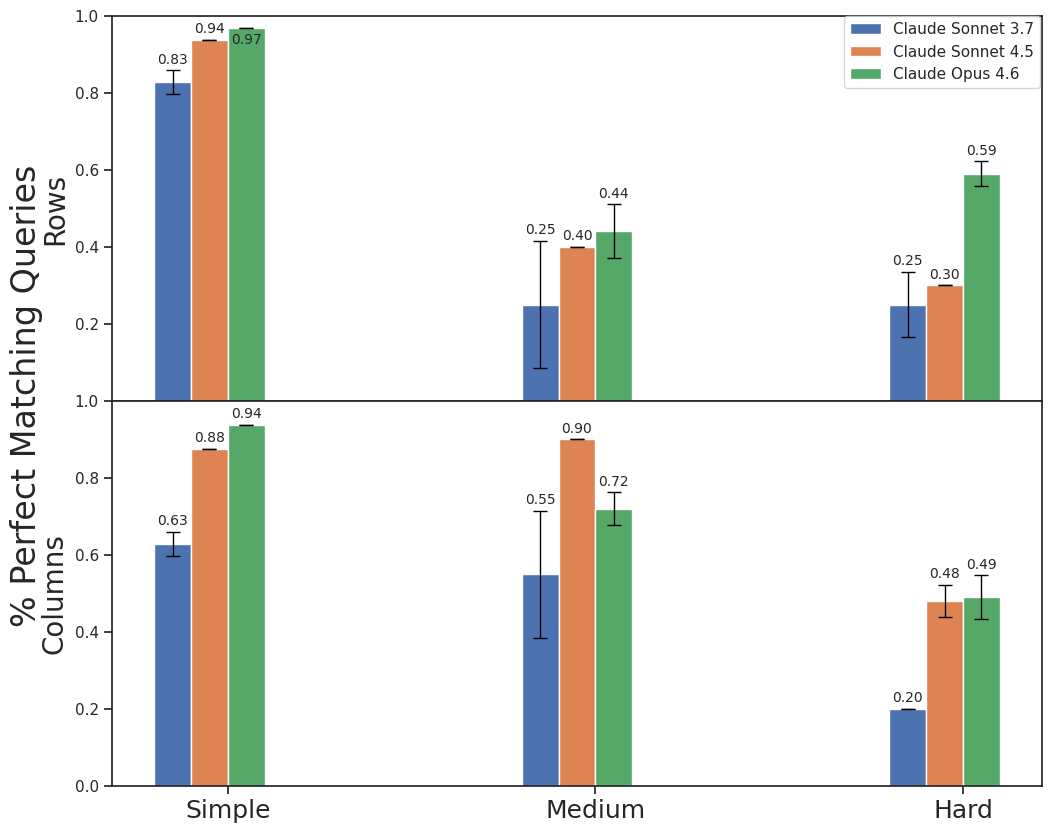}
        \caption{Step-by-step generation method}
    \end{subfigure}%
    \caption{Performance comparison of Claude models, as a function of the level of query difficulty.}
    \label{fig:claude_llm_comparison}
\end{figure}

\begin{figure}[h!]
    \centering
    \begin{subfigure}[b]{0.4\textwidth}
        \centering
        \includegraphics[width=1.0\linewidth]{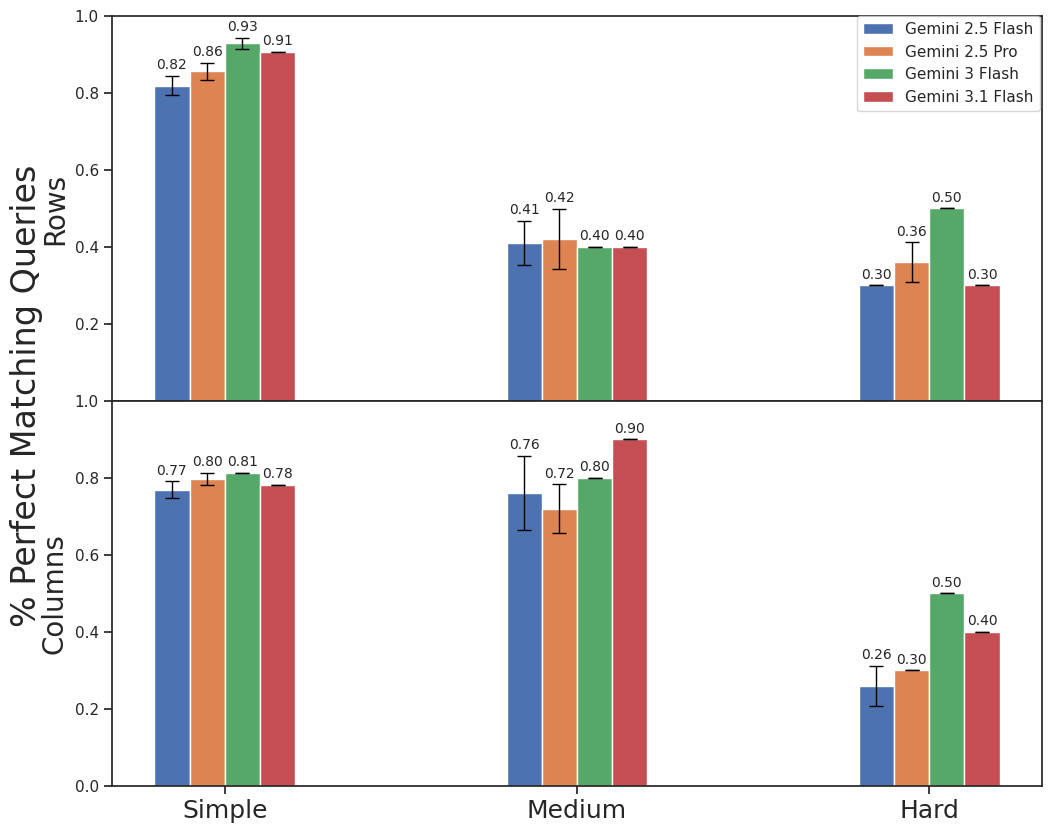}
        \caption{Direct generation method}
    \end{subfigure}%
    \begin{subfigure}[b]{0.4\textwidth}
        \centering
      \includegraphics[width=1.0\linewidth]{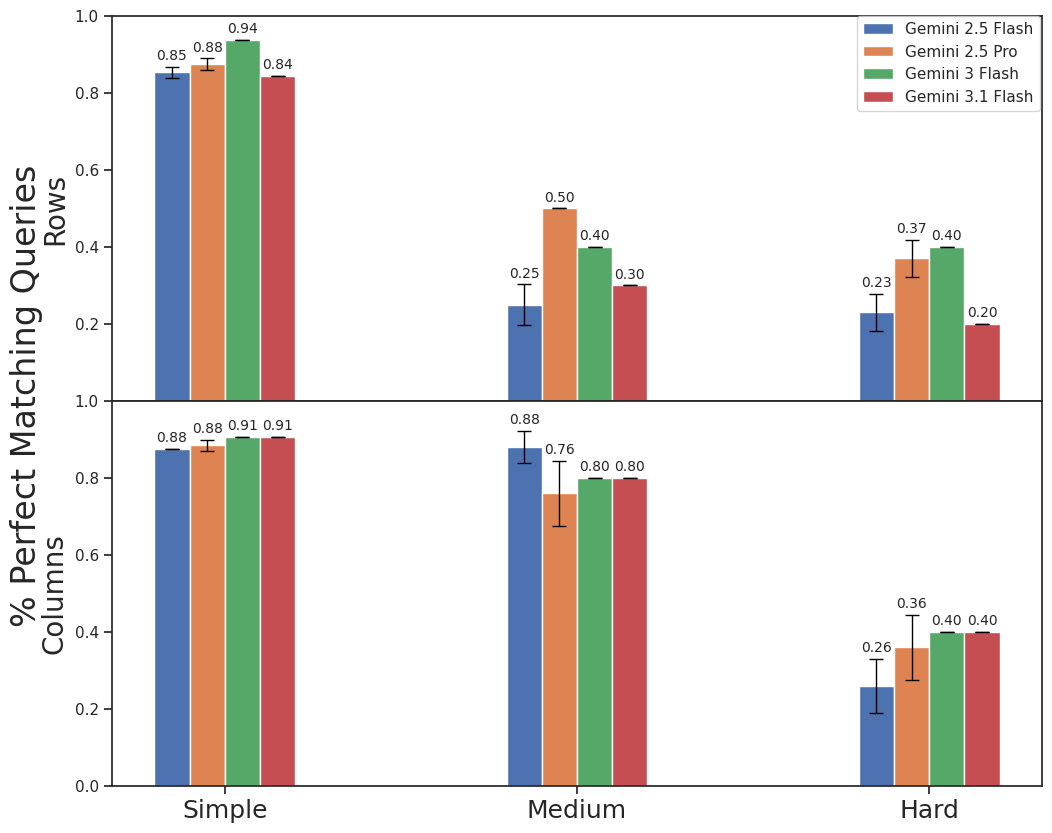}
        \caption{Step-by-step generation method}
    \end{subfigure}%
    \caption{Performance comparison of Gemini models, as a function of the level of query difficulty.}
    \label{fig:gemini_llm_comparison}
\end{figure}

\begin{figure}[h!]
    \centering
    \begin{subfigure}[b]{0.4\textwidth}
        \centering
        \includegraphics[width=1.0\linewidth]{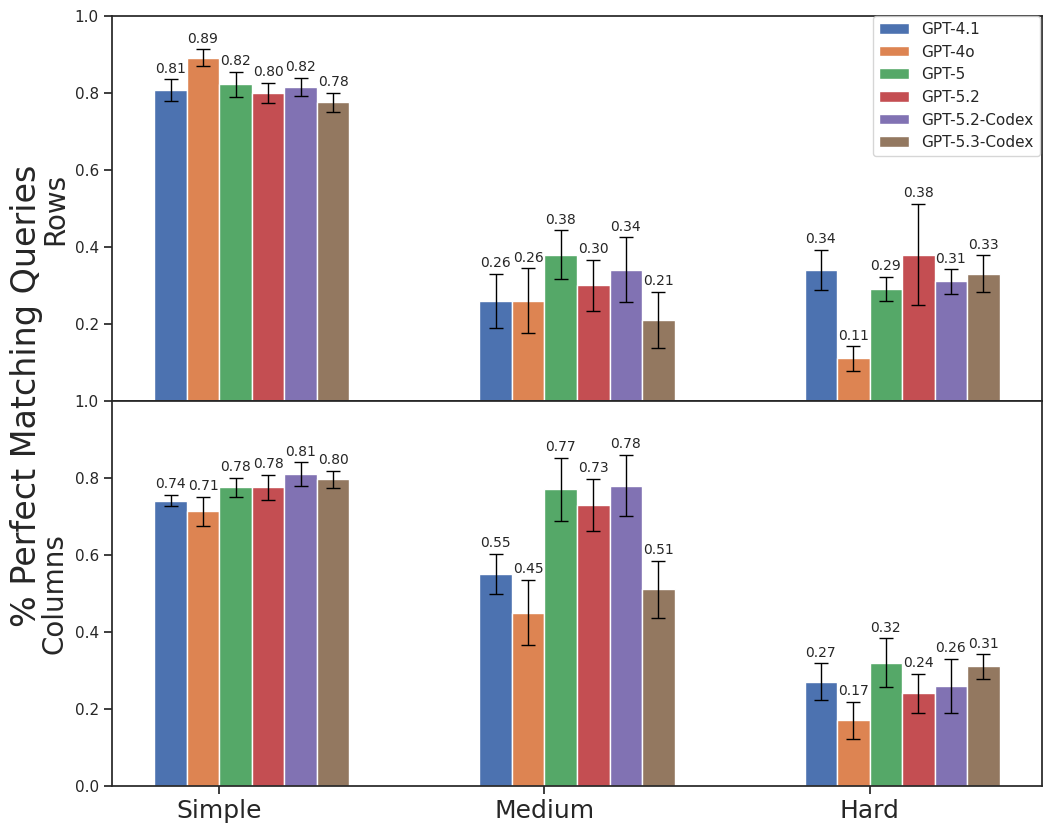}
        \caption{Direct generation method}
    \end{subfigure}%
    \begin{subfigure}[b]{0.4\textwidth}
        \centering
      \includegraphics[width=1.0\linewidth]{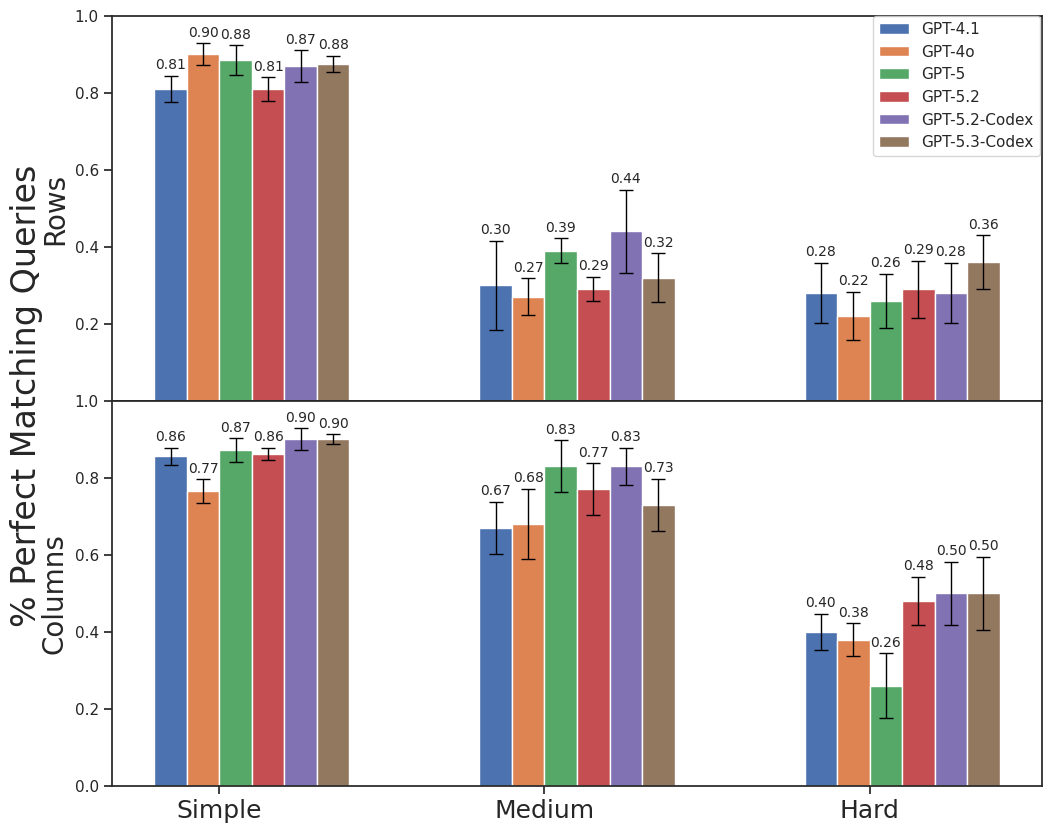}
        \caption{Step-by-step generation method}
    \end{subfigure}%
    \caption{Performance comparison of GPT models, as a function of the level of query difficulty.}
    \label{fig:gpt_llm_comparison}
\end{figure}

\begin{figure}[h!]
     \sidecaption
     \includegraphics[width=0.5\linewidth]{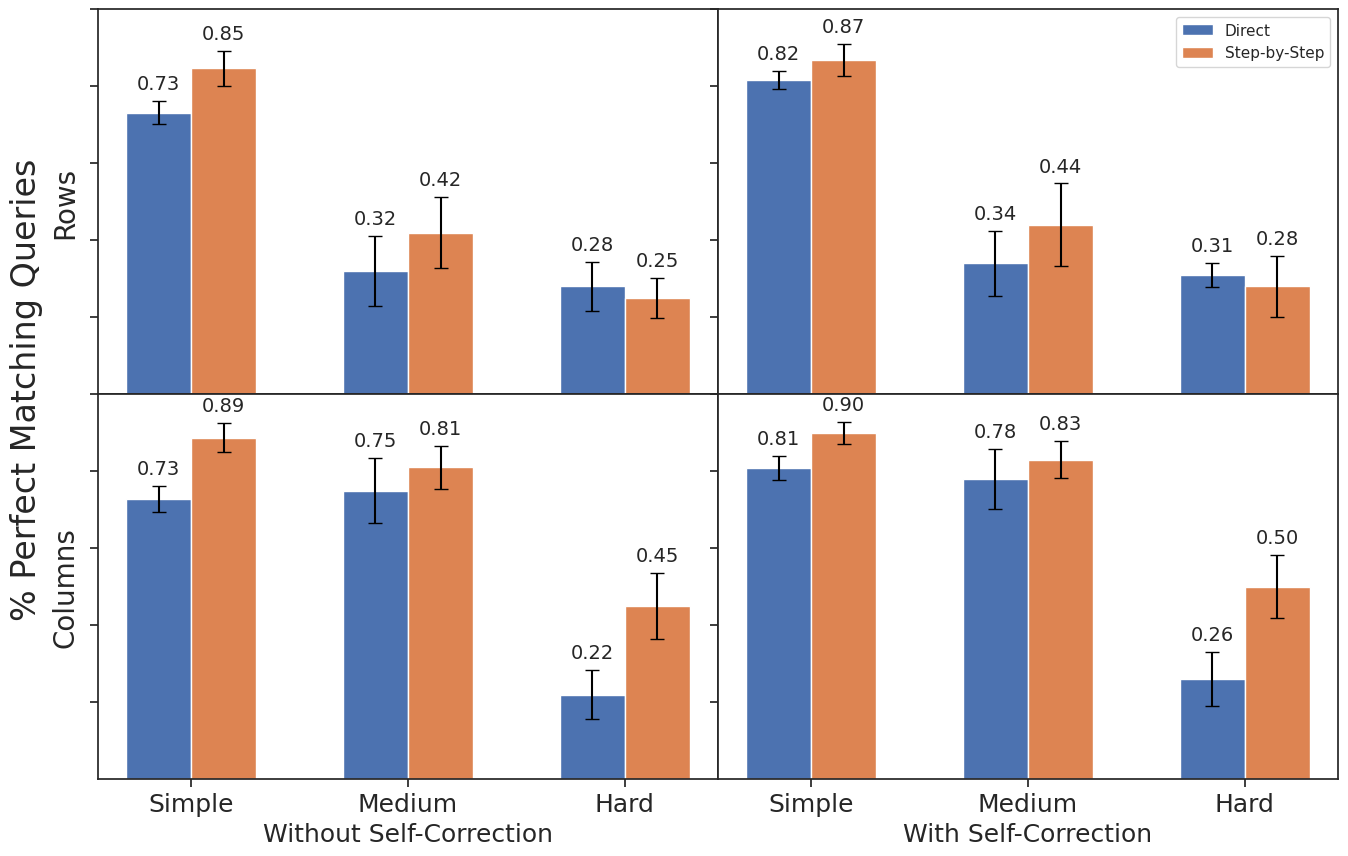}
    \caption{GPT-5.2-Codex performance without vs with self correction for direct and step-by-step generation methods.}
    \label{fig:nosc_vs_sc_gpt52codex}
\end{figure}

\begin{figure}[h!]
     \sidecaption
     \includegraphics[width=0.6\linewidth]{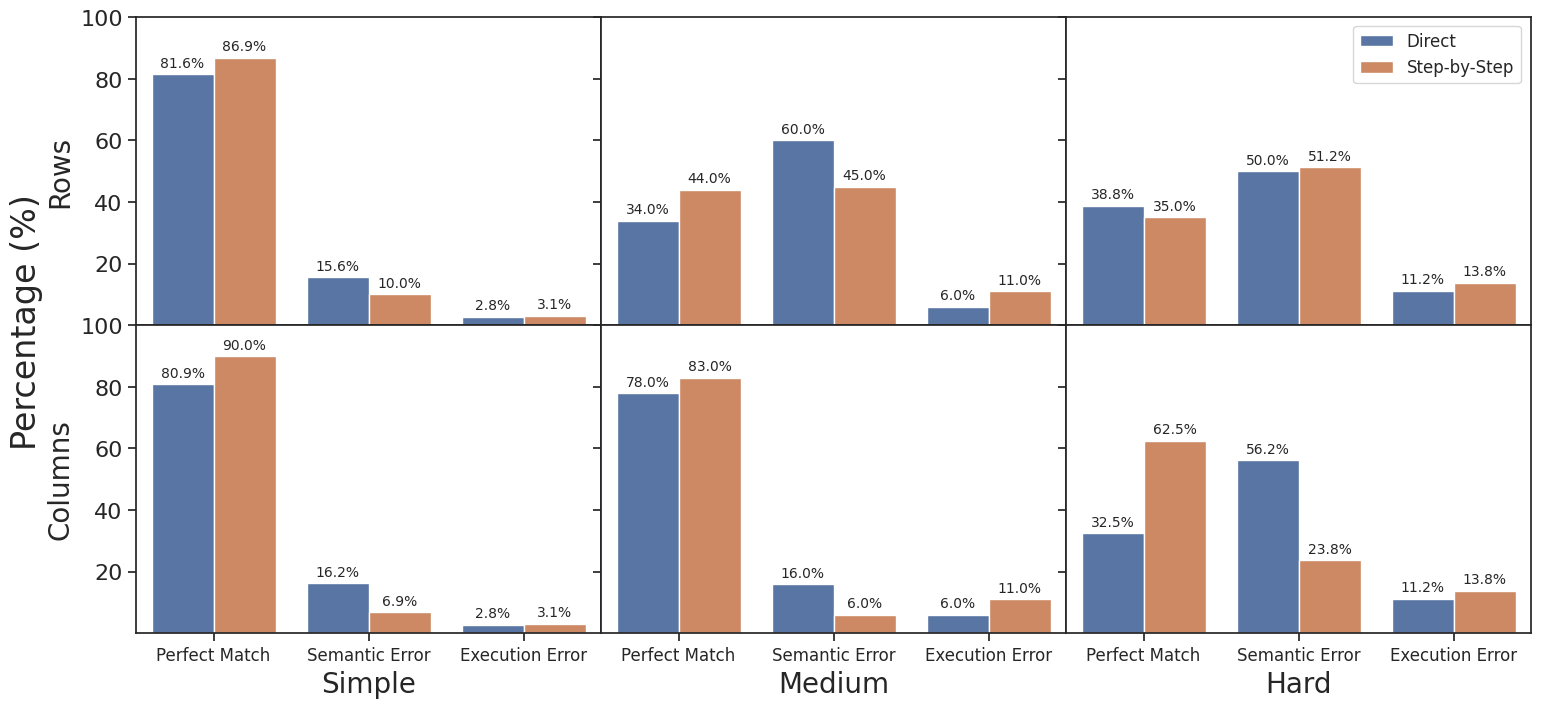}
    \caption{Percentage of perfect queries, semantic errors, and execution errors for the direct and step-by-step query generation methods using GPT-5.2-Codex.}
    \label{fig:percentage_errors_dir_vs_sbs_gpt52codex}
\end{figure}

\begin{figure}[H]
     \sidecaption
     \includegraphics[width=0.6\linewidth]{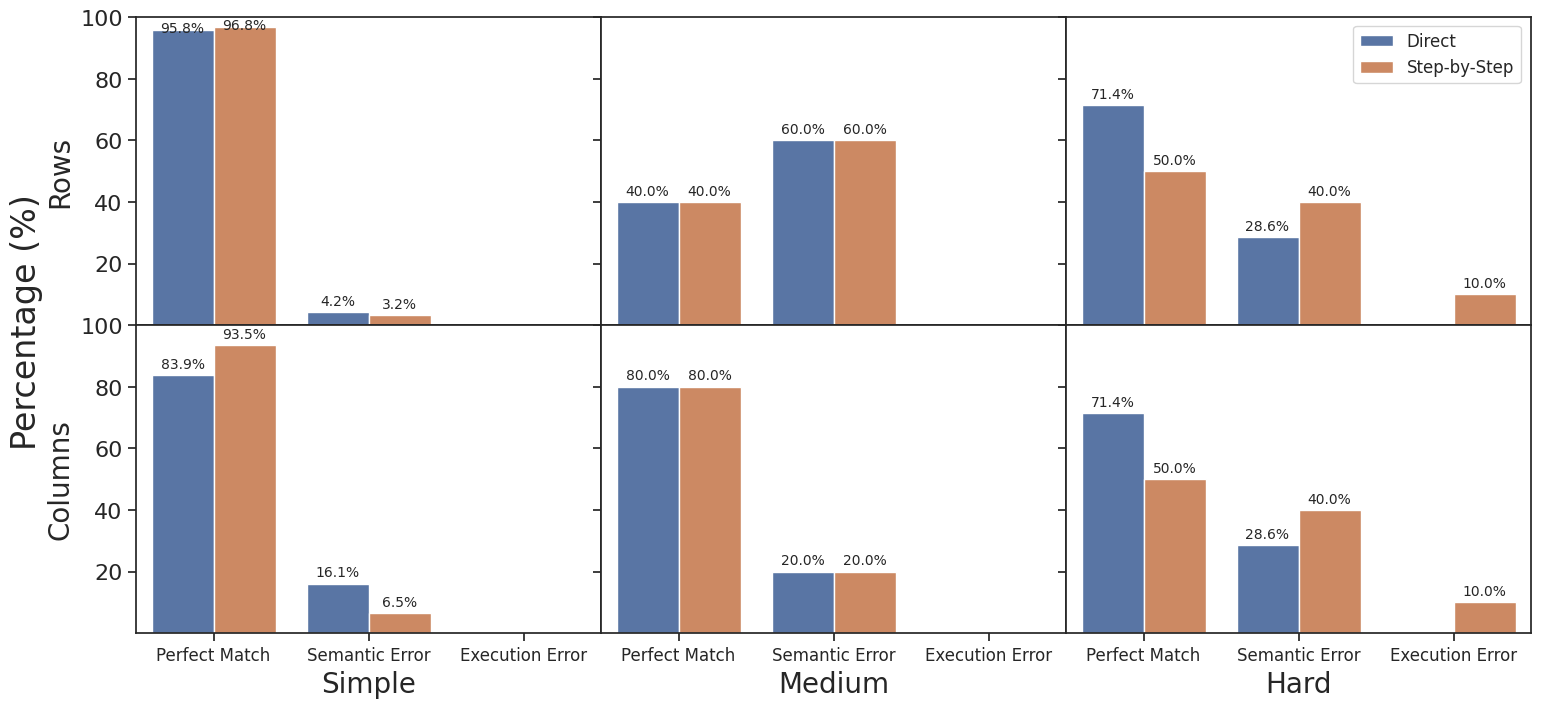}
    \caption{Number of perfect queries, semantic errors, and execution errors for the direct and step-by-step query generation methods using Gemini 3 Flash.}
    \label{fig:count_errors_dir_vs_sbs_gemini3flash}
\end{figure}

\section{Few-shot prompting}
\label{appendix:few_shot}

\renewcommand{\thelstlisting}{F.\arabic{lstlisting}}

\begin{table}[htbp!]
\footnotesize
\centering
\caption{Few-shot PM results by query difficulty when using Gemini 2.5 Pro (step-by-step).}
\label{tab:few_shot_comparison_by_difficulty_sbs_gemini25pro}
\begin{tabular}{llcccccc}
\toprule
\toprule
Few-shot & Selection & \multicolumn{6}{c}{Gemini 2.5 Pro} \\
\cmidrule(lr){3-8}
 &  & \multicolumn{2}{c}{Simple} & \multicolumn{2}{c}{Medium} & \multicolumn{2}{c}{Hard} \\
\cmidrule(lr){3-4} \cmidrule(lr){5-6} \cmidrule(lr){7-8}
 &  & Rows & Columns & Rows & Columns & Rows & Columns \\
\midrule
0-shot & - & 0.88 $\pm$ 0.015 & 0.88 $\pm$ 0.015 & 0.50 $\pm$ 0.000 & 0.76 $\pm$ 0.084 & 0.37 $\pm$ 0.048 & 0.36 $\pm$ 0.084 \\
\midrule
1-shot & Random & \textbf{0.97 $\pm$ 0.000 $\uparrow$} & 0.92 $\pm$ 0.017 $\sim$ & 0.50 $\pm$ 0.000 $\sim$ & 0.82 $\pm$ 0.084 $\sim$ & 0.32 $\pm$ 0.045 $\downarrow$ & 0.40 $\pm$ 0.122 $\sim$ \\
 & QTS & 0.96 $\pm$ 0.017 $\uparrow$ & 0.94 $\pm$ 0.014 $\uparrow$ & 0.50 $\pm$ 0.000 $\sim$ & 0.74 $\pm$ 0.055 $\sim$ & 0.36 $\pm$ 0.114 $\sim$ & 0.38 $\pm$ 0.084 $\sim$ \\
 & MQS & 0.95 $\pm$ 0.017 $\uparrow$ & 0.93 $\pm$ 0.017 $\uparrow$ & 0.50 $\pm$ 0.000 $\sim$ & 0.82 $\pm$ 0.045 $\sim$ & 0.34 $\pm$ 0.055 $\sim$ & \textbf{0.46 $\pm$ 0.055 $\sim$} \\
\midrule
3-shot & Random & 0.95 $\pm$ 0.028 $\uparrow$ & 0.94 $\pm$ 0.000 $\uparrow$ & \textbf{0.60 $\pm$ 0.000 $\uparrow$} & 0.82 $\pm$ 0.045 $\sim$ & 0.40 $\pm$ 0.122 $\sim$ & 0.42 $\pm$ 0.045 $\sim$ \\
 & QTS & 0.95 $\pm$ 0.017 $\uparrow$ & \textbf{0.95 $\pm$ 0.017 $\uparrow$} & 0.50 $\pm$ 0.000 $\sim$ & \textbf{0.84 $\pm$ 0.055 $\sim$} & \textbf{0.48 $\pm$ 0.045 $\uparrow$} & \textbf{0.46 $\pm$ 0.055 $\uparrow$} \\
 & MQS & 0.94 $\pm$ 0.014 $\uparrow$ & 0.94 $\pm$ 0.000 $\uparrow$ & 0.50 $\pm$ 0.000 $\sim$ & 0.82 $\pm$ 0.045 $\sim$ & 0.32 $\pm$ 0.045 $\downarrow$ & 0.36 $\pm$ 0.055 $\sim$ \\
\bottomrule
\end{tabular}
\tablefoot{The best value per table column is bolded. Arrows compare the few-shot setting against the 0-shot baseline for each query difficulty using two-sided permutation tests on perfect-match scores grouped by run; $\uparrow$/$\downarrow$ indicate significantly higher/lower results at $\alpha=0.05$ with no multiple-testing correction, and $\sim$ indicates not significant.}
\end{table}

\begin{table}[htbp!]
\footnotesize
\centering
\caption{Few-shot PM results by query difficulty when using Claude Opus 4.6 (step-by-step).}
\label{tab:few_shot_comparison_by_difficulty_sbs_claudeopus46}
\begin{tabular}{llcccccc}
\toprule
Few-shot & Selection & \multicolumn{6}{c}{Claude Opus 4.6} \\
\cmidrule(lr){3-8}
 &  & \multicolumn{2}{c}{Simple} & \multicolumn{2}{c}{Medium} & \multicolumn{2}{c}{Hard} \\
\cmidrule(lr){3-4} \cmidrule(lr){5-6} \cmidrule(lr){7-8}
 &  & Rows & Columns & Rows & Columns & Rows & Columns \\
\midrule
0-shot & - & 0.97 $\pm$ 0.000 & 0.94 $\pm$ 0.000 & \textbf{0.44 $\pm$ 0.070} & 0.72 $\pm$ 0.042 & \textbf{0.59 $\pm$ 0.032} & 0.49 $\pm$ 0.057 \\
\midrule
1-shot & Random & 0.97 $\pm$ 0.000 $\sim$ & 0.93 $\pm$ 0.017 $\sim$ & 0.32 $\pm$ 0.045 $\downarrow$ & 0.72 $\pm$ 0.045 $\sim$ & 0.42 $\pm$ 0.045 $\downarrow$ & 0.46 $\pm$ 0.055 $\sim$ \\
 & QTS & 0.97 $\pm$ 0.000 $\sim$ & 0.94 $\pm$ 0.000 $\sim$ & 0.30 $\pm$ 0.000 $\downarrow$ & 0.70 $\pm$ 0.000 $\sim$ & 0.48 $\pm$ 0.045 $\downarrow$ & \textbf{0.50 $\pm$ 0.000 $\sim$} \\
 & MQS & 0.97 $\pm$ 0.000 $\sim$ & 0.96 $\pm$ 0.017 $\sim$ & 0.42 $\pm$ 0.045 $\sim$ & \textbf{0.82 $\pm$ 0.045 $\uparrow$} & 0.52 $\pm$ 0.045 $\downarrow$ & 0.48 $\pm$ 0.045 $\sim$ \\
\midrule
3-shot & Random & 0.90 $\pm$ 0.014 $\downarrow$ & 0.88 $\pm$ 0.014 $\downarrow$ & 0.38 $\pm$ 0.084 $\sim$ & \textbf{0.82 $\pm$ 0.045 $\uparrow$} & 0.38 $\pm$ 0.045 $\downarrow$ & 0.42 $\pm$ 0.084 $\sim$ \\
 & QTS & \textbf{0.97 $\pm$ 0.014 $\sim$} & \textbf{0.97 $\pm$ 0.000 $\uparrow$} & 0.40 $\pm$ 0.000 $\sim$ & 0.70 $\pm$ 0.000 $\sim$ & 0.48 $\pm$ 0.045 $\downarrow$ & 0.48 $\pm$ 0.045 $\sim$ \\
 & MQS & 0.94 $\pm$ 0.000 $\downarrow$ & 0.94 $\pm$ 0.000 $\sim$ & 0.30 $\pm$ 0.000 $\downarrow$ & 0.70 $\pm$ 0.000 $\sim$ & 0.58 $\pm$ 0.045 $\sim$ & 0.46 $\pm$ 0.055 $\sim$ \\
\bottomrule
\end{tabular}
\tablefoot{The best value per table column is bolded. Arrows compare the few-shot setting against the 0-shot baseline for each query difficulty using two-sided permutation tests on perfect-match scores grouped by run; $\uparrow$/$\downarrow$ indicate significantly higher/lower results at $\alpha=0.05$ with no multiple-testing correction, and $\sim$ indicates not significant.}
\end{table}

\begin{lstlisting}[language=python,  caption={Example few-shot prompt}, label={lst:fewshot_example} ,basicstyle=\ttfamily\scriptsize, float=htbp!]
## Example 1 (tables: dataquality, ...)
# Important Information for the query
External Knowledge:
# User Request: ...
```sql
...
```
## End Example

# Important Information for the query
External Knowledge:
# User Request:
\end{lstlisting}

\clearpage
\section{Function calling and structured outputs}
\label{appendix:func_call}

\renewcommand{\thelstlisting}{G.\arabic{lstlisting}}

\begin{table}[htbp!]
\footnotesize
\centering
\setlength{\tabcolsep}{4pt}
\caption{Perfect match rates by LLM, query generation method, query difficulty, and w/ or w/o function calling (w/o self-correction).}
\label{tab:combined_vertical_results_by_function_calling}
\begin{tabular}{lcccccc}
\toprule
\toprule
 & \multicolumn{2}{c}{Simple} & \multicolumn{2}{c}{Medium} & \multicolumn{2}{c}{Hard} \\
\cmidrule(lr){2-3} \cmidrule(lr){4-5} \cmidrule(lr){6-7}
Model & w/o Func. call & w/ Func. call & w/o Func. call & w/ Func. call & w/o Func. call & w/ Func. call \\
\midrule
\multicolumn{7}{l}{\textbf{ID Match (Rows) -- Direct}} \\
Claude Opus 4.6 & 0.844 $\pm$ 0.000 & \textbf{0.850 $\pm$ 0.014} & 0.250 $\pm$ 0.053 & 0.280 $\pm$ 0.045 & \textbf{0.400 $\pm$ 0.000} & \textbf{0.460 $\pm$ 0.055} \\
Gemini 2.5 Pro & 0.809 $\pm$ 0.023 & 0.787 $\pm$ 0.014 & \textbf{0.360 $\pm$ 0.117} & \textbf{0.320 $\pm$ 0.045} & 0.160 $\pm$ 0.070 & 0.200 $\pm$ 0.071 \\
Gemini 3 Flash & \textbf{0.866 $\pm$ 0.015} & 0.769 $\pm$ 0.034 & 0.300 $\pm$ 0.000 & \textbf{0.320 $\pm$ 0.042} & 0.300 $\pm$ 0.000 & 0.360 $\pm$ 0.052 \\
\cmidrule(lr){1-7}
\multicolumn{7}{l}{\textbf{ID Match (Rows) -- Step-by-Step}} \\
Claude Opus 4.6 & 0.938 $\pm$ 0.000 & 0.938 $\pm$ 0.000 & 0.380 $\pm$ 0.042 & 0.390 $\pm$ 0.088 & \textbf{0.530 $\pm$ 0.048} & \textbf{0.460 $\pm$ 0.084} \\
Gemini 2.5 Pro & 0.847 $\pm$ 0.023 & 0.863 $\pm$ 0.017 & \textbf{0.490 $\pm$ 0.032} & \textbf{0.500 $\pm$ 0.000} & 0.250 $\pm$ 0.053 & 0.240 $\pm$ 0.055 \\
Gemini 3 Flash & \textbf{0.969 $\pm$ 0.000} & \textbf{0.944 $\pm$ 0.025} & 0.290 $\pm$ 0.032 & 0.310 $\pm$ 0.088 & 0.300 $\pm$ 0.000 & 0.240 $\pm$ 0.052 \\
\midrule
\multicolumn{7}{l}{\textbf{Column Match -- Direct}} \\
Claude Opus 4.6 & 0.797 $\pm$ 0.016 & 0.787 $\pm$ 0.014 & 0.640 $\pm$ 0.052 & 0.600 $\pm$ 0.071 & 0.320 $\pm$ 0.042 & 0.300 $\pm$ 0.000 \\
Gemini 2.5 Pro & 0.797 $\pm$ 0.016 & \textbf{0.831 $\pm$ 0.017} & 0.660 $\pm$ 0.084 & 0.560 $\pm$ 0.089 & 0.180 $\pm$ 0.079 & 0.220 $\pm$ 0.045 \\
Gemini 3 Flash & \textbf{0.803 $\pm$ 0.015} & 0.806 $\pm$ 0.020 & \textbf{0.800 $\pm$ 0.000} & \textbf{0.710 $\pm$ 0.057} & \textbf{0.400 $\pm$ 0.000} & \textbf{0.420 $\pm$ 0.042} \\
\cmidrule(lr){1-7}
\multicolumn{7}{l}{\textbf{Column Match -- Step-by-Step}} \\
Claude Opus 4.6 & \textbf{0.938 $\pm$ 0.000} & \textbf{0.938 $\pm$ 0.000} & 0.650 $\pm$ 0.053 & 0.660 $\pm$ 0.126 & \textbf{0.480 $\pm$ 0.042} & \textbf{0.460 $\pm$ 0.052} \\
Gemini 2.5 Pro & 0.884 $\pm$ 0.015 & 0.869 $\pm$ 0.026 & 0.750 $\pm$ 0.071 & 0.780 $\pm$ 0.084 & 0.290 $\pm$ 0.057 & 0.260 $\pm$ 0.055 \\
Gemini 3 Flash & \textbf{0.938 $\pm$ 0.000} & \textbf{0.938 $\pm$ 0.000} & \textbf{0.800 $\pm$ 0.000} & \textbf{0.830 $\pm$ 0.095} & 0.400 $\pm$ 0.000 & 0.420 $\pm$ 0.042 \\
\bottomrule
\end{tabular}
\tablefoot{Best results per column in each of the four sections of the table are bolded.}
\end{table}

\begin{table}[htbp!]
\footnotesize
\centering
\setlength{\tabcolsep}{4pt}
\caption{Structured vs.\ base step-by-step plan comparison, by difficulty (with self-correction).}
\label{tab:structured_plan_comparison_by_difficulty}
\begin{tabular}{lcccc}
\toprule
\toprule
 & \multicolumn{2}{c}{Medium} & \multicolumn{2}{c}{Hard} \\
\cmidrule(lr){2-3} \cmidrule(lr){4-5}
Model & Free-form & Structured outputs & Free-form & Structured outputs \\
\midrule
\multicolumn{5}{l}{\textbf{Rows}} \\
Gemini 3 Flash & 0.40 $\pm$ 0.000 & 0.30 $\pm$ 0.100 $\downarrow$ & 0.40 $\pm$ 0.000 & 0.38 $\pm$ 0.045 $\sim$ \\
Gemini 2.5 Pro & \textbf{0.50 $\pm$ 0.000} & \textbf{0.50 $\pm$ 0.000} & 0.37 $\pm$ 0.048 & 0.34 $\pm$ 0.055 $\sim$ \\
Claude Opus 4.6 & 0.44 $\pm$ 0.070 & 0.30 $\pm$ 0.000 $\downarrow$ & \textbf{0.59 $\pm$ 0.032} & \textbf{0.58 $\pm$ 0.130} $\sim$ \\
\midrule
\multicolumn{5}{l}{\textbf{Columns}} \\
Gemini 3 Flash & \textbf{0.81 $\pm$ 0.032} & \textbf{0.88 $\pm$ 0.045} $\uparrow$ & 0.30 $\pm$ 0.000 & \textbf{0.30 $\pm$ 0.000} \\
Gemini 2.5 Pro & 0.80 $\pm$ 0.047 & 0.84 $\pm$ 0.055 $\sim$ & \textbf{0.31 $\pm$ 0.057} & 0.28 $\pm$ 0.045 $\sim$ \\
Claude Opus 4.6 & 0.72 $\pm$ 0.042 & 0.70 $\pm$ 0.000 $\sim$ & \textbf{0.31 $\pm$ 0.032} & 0.28 $\pm$ 0.045 $\sim$ \\
\bottomrule
\end{tabular}
\tablefoot{Each pair shows rows and columns as mean $\pm$ std over run-level means; best per column is bolded. Markers on structured cells reflect a unpaired permutation test, $\uparrow$/$\downarrow$ when the adjusted $p$-value rejects equality, $\sim$ otherwise.}
\end{table}

\begin{lstlisting}[language=Python,  caption={Specialized tool defined in Python to transform date formats to MJD.}, label={lst:specialized_tool_example} ,basicstyle=\ttfamily\scriptsize, float=htbp!]
date_to_mjd_tool = types.FunctionDeclaration(
            name="date_to_mjd",
            description=(
                "Convert ONE date expression to a single MJD value. Accepts ISO 8601 (e.g. '2022-12-01'), or relative expressions anchored to the current time (e.g. '300 days ago', '7 days ago'). "
                "For date ranges, call this tool once per boundary. "
                "ALeRCE stores all dates as MJD."
            ),
            parameters=types.Schema(
                type=types.Type.OBJECT,
                properties={
                    "date_expression": types.Schema(
                        type=types.Type.STRING,
                        description="A single date: ISO 8601 ('2023-09-01') or relative ('300 days ago', '7 days ago')."
                    ),
                    "label": types.Schema(
                        type=types.Type.STRING,
                        description="A short snake_case key summarizing the date as mjd_[month]_[day] or mjd_[relative]_[count], e.g. 'mjd_august_23' or 'mjd_hours_1'."
                    ),
                },
                required=["date_expression"],
            ),
        )
\end{lstlisting}

\begin{lstlisting}[language=Python,  caption={Pydantic model for the structured output step-by-step planning.}, label={lst:sbs_plan_pydantic} ,basicstyle=\ttfamily\scriptsize, float=htbp!]

class PlanStep(BaseModel):
    step: str = Field(
        description="A single decomposition step describing part of the plan to generate the SQL query.")
class DecompositionPlanOutput(BaseModel):
    steps: list[PlanStep] = Field(
        description="Ordered decomposition steps for the SQL generation plan.",
        min_length=1,)
\end{lstlisting}

\begin{lstlisting}[language=Python,  caption={Pydantic format converted to JSON schema for API calls.}, label={lst:sbs_plan_json} ,basicstyle=\ttfamily\scriptsize, float=htbp!]

{"type": "json_schema",
    "name": "DecompositionPlanOutput",
    "strict": true,
    "schema": {
      "$defs": {
        "PlanStep": {
          "additionalProperties": false,
          "properties": {
            "step": {
              "description": "A single decomposition step describing part of the plan to generate the SQL query.",
              "title": "Step",
              "type": "string"
            }
          },
          "required": [
            "step"
          ],
          "title": "PlanStep",
          "type": "object"
        }
      },
      "additionalProperties": false,
      "properties": {
        "steps": {
          "description": "Ordered decomposition steps for the SQL generation plan.",
          "items": {
            "$ref": "#/$defs/PlanStep"
          },
          "minItems": 1,
          "title": "Steps",
          "type": "array"
        }
      },
      "required": [
        "steps"
      ],
      "title": "DecompositionPlanOutput",
      "type": "object"
    }
  }
    
\end{lstlisting}

\end{appendix}

\end{document}